# Multi-objective Bayesian optimization of ferroelectric materials with interfacial control for memory and energy storage applications

Arpan Biswas,[1] Anna N. Morozovska,[2,a] Maxim Ziatdinov,[1,4]

Eugene A. Eliseev,[3] and Sergei V. Kalinin[1,b]

[1] Center for Nanophase Materials Sciences, Oak Ridge National Laboratory, Oak Ridge, TN 37831

[2] Institute of Physics, National Academy of Sciences of Ukraine, pr. Nauky 46, 03028 Kyiv, Ukraine

[3] Institute for Problems of Materials Science, National Academy of Sciences of Ukraine, Krjijanovskogo 3, 03142 Kyiv, Ukraine

[4] Computational Sciences and Engineering Division, Oak Ridge National Laboratory, Oak Ridge, TN 37831

Optimization of materials performance for specific applications often requires balancing multiple aspects of materials functionality. Even for the cases where generative physical model of material behavior is known and reliable, this often requires search over multidimensional parameter space to identify low-dimensional manifold corresponding to required Pareto front. Here we introduce the multi-objective Bayesian Optimization (MOBO) workflow for the ferroelectric/anti-ferroelectric performance optimization for memory and energy storage applications based on the numerical solution of the Ginzburg-Landau equation with electrochemical or semiconducting boundary conditions. MOBO is a low computational cost optimization tool for expensive multi-objective functions, where we update posterior surrogate Gaussian process models from prior evaluations, and then select future evaluations from maximizing an acquisition function. Using the parameters for a prototype bulk antiferroelectric (PbZrO$_3$), we first develop a physics-driven decision tree of target functions from the loop structures. We further develop a physics-driven MOBO architecture to explore multidimensional parameter space and build Pareto-frontiers by maximizing two target functions jointly- energy storage and loss. This approach allows for rapid initial materials and device parameter selection for a given application and can be further expanded towards the active experiment setting. The associated notebooks provide both the tutorial on MOBO and allow to reproduce the reported analyses and apply them to other systems (https://github.com/arpanbiswas52/MOBO_AFI_Supplements).

[a] anna.n.morozovska@gmail.com

[b] sergei2@ornl.gov

Ferroelectric and antiferroelectric materials are one of the key materials classes for the memory and energy storage applications.[1] For ferroelectrics, the classical applications include ferroelectric capacitors,[2,3] as well as emergent applications such as ferroelectric tunneling devices[4–6] and long-sought ferroelectric field effect transistors.[7,8] The complex ferroelectric structures are of broad interest in the context of ferroelectric device integration, as well as enabling the new functionalities such as composite magnetoelectric devices.[9,10] Similarly, antiferroelectric materials are broadly considered as potential energy storage, either in the form of the on-chip components, or in the capacitive or electrolytically gated configurations. Common for both ferroelectric and antiferroelectric materials is the presence of polar instability, which is a ground state for ferroelectric and is suppressed by the structural instability at zero fields for antiferroelectrics.[11,12] Correspondingly, FE and AFE materials both allow for description in the form of the Ginzburg-Landau-Devonshire (GLD) type theories[1,13–15] with several competing order parameters, which include a polarization for multiaxial FE and a structural order parameter for the AFE systems. This formalism allows for direct prediction of corresponding ground states, concentration, field and strain responses, as well as serves as the basis for the phase-field models.[16–18]

The potential applications of ferroelectrics for information technology have stimulate over two decades of effort in integrating the classical perovskite-based systems with semiconductors, stymied by the intrinsic instability of the oxide-semiconductor interfaces. The development of $HfO_2$ and other binary ferroelectrics by Mikolajick $et. al.$ in early 2011 have stimulated new and rapidly growing research effort in this area,[19–23] also stimulating the search for and discovery of novel binary systems such as $(Al_xSc_{1-x}N)$[24,25] and $(Mg_{1-x}Zn_xO)$[26] Notably, despite significant controversies on origin of ferroelectric-like responses in these systems,[27–31] these materials systems also allow effective representation via GLD formalism, predicting dopant-dependent phase transitions and temperature-polarization behavior.[32,33]

Both for conventional and binary ferroelectrics, of fundamental interest is their behavior and performance in the realistic devices and structures, i.e. in the form of films, multilayers, or 3D integrated objects.[34–41] The intrinsic complexity of this problem stems from the non-local nature of the depolarization phenomena in ferroelectric systems, where the distributions of the polarization and screening charges on the bounding interfaces result in the dominant contribution to the thermodynamics of the system. Traditionally, these phenomena are analyzed using effective dead layer models, where the complex electrostatic structure of interface is approximated via introduction of effective dielectric layer.[14,42] However, this approximation is limited to more realistic scenarios including open ferroelectric surface with ionic screening, interfaces between ferroelectric and ionically conductive and electrochemically active materials, and interfaces between combining ferroelectric and semiconductor functionalities, i.e. ferroelectrics-semiconductor interfaces,[43,44] interfaces between ferroelectric-semiconductors and other materials, and interfaces with intrinsic finite density of electronic (or ionic) states.

Despite apparent variability of these systems, this broad range of phenomena allows ready representation in terms of formalism combining the LGD description of ferroelectric system with the corresponding surface and bulk semiconductor subsystem, or surface/interface electrochemical models. For many ferroelectric materials corresponding free energy coefficients are well known, as are their temperature, concentration, and strain dependencies, whereas other can be readily measured via classical semiconductor characterization techniques. As such, they are amenable for numerical assessment via approximate analytical, 1D numerical, and 3D numerical models.

However, many of these models are time consuming, and bring forth the challenge of effective navigation of possible parameter space towards target functionalities or their combination.

Here, we introduce the combination of Multi Objective Bayesian Optimization (MOBO) and ferroelectric-electrochemical model for figure of merit optimization for memory and energy storage devices. The analysis here is developed for 1D model that does not allow for formation of domain structures, but can be trivially extended to the full 3D finite element models. We develop the general formalism for the (anti)ferroelectric with the interface control in Section I, introduce the Bayesian Optimization and Multi Objective Bayesian Optimization in Section II, and demonstrate the exploration of optimal functionalities in the interface-controlled ferroelectrics in Section III. The analyses reported here are summarized in the associated Colab notebooks that can both serve as tutorials and allow to reproduce the results in this manuscript and apply the same analytics for other models. (https://github.com/arpanbiswas52/MOBO_AFI_Supplements).

## 1. General ferroelectric – ionic/semiconductor model

Traditionally, interfacial phenomena in ferroelectrics are studied in the dielectric dead layer approximation, effectively allowing for separation between the polarization bound charges and screening charges. While capturing some aspects of polarization behavior in ferroelectrics, this model is insufficient to describe the surface and interface phenomena in ferroelectrics in the presence of finite electronic or ionic density of states. For ferroelectric-semiconductor interfaces nonlinear screening phenomena and their influence on the contact barrier can be important, while the phenomena are usually either neglected (model of constant screening charge density) or linear approximation is used (Bardeen states model). [45–47] Similarly, a number of authors have explored ferroelectric semiconductors, where both ferroelectric and semiconductive subsystems coexist in a single material.[48] Finally, for surfaces complementary thermodynamic approach was developed by Stephenson and Highland (SH)[49,50], who consider an ultrathin film in equilibrium with a chemical environment that supplies (positive and negative) ionic species to compensate its polarization bound charge at the surface.

Recently, we modified the SH approach allowing for the presence of a gap between the ferroelectric surface covered by ions and a SPM tip,[39,51–54] and developed the analytical description for thermodynamics and kinetics of these systems. The analysis leads to the elucidation of ferroionic states, which are the result of nonlinear electrostatic interaction between the ions with the surface charge density described via Langmuir adsorption isotherm and ferroelectric dipoles. Recently, the emergent behaviors in uniaxial antiferroelectric (AFE) thin films covered with surface ions have been explored within the framework of the Kittel-Landau-Ginzburg-Devonshire-Stephenson-Highland (KLGD-SH) approach.[55] The approach allows the characterization of the polar and antipolar orderings dependence on the voltage applied to the antiferroelectric film, and the analysis of their static and dynamic hysteresis loops.

Below we introduce the general formalism for polarization rotation in the AFE films covered by surface ions with a charge density proportional to the relative partial oxygen pressure. We calculate the phase diagrams and analyze the dependence of polarization and anti-polarization on applied voltage, and discuss the peculiarities of static and dynamic hysteresis loops. It is important to note that, while here we consider a special case of adsorption isotherm, this analysis can be readily extended towards other functional forms and mechanisms that give rise to finite electron or ion density of states at the interface. While the exact nature of the carries will affect the transport and reaction mechanisms, the ferroelectric responses are expected to be universal.

Here we consider the case of ultra-thin dielectric gap between the top electrode and the surface of AFE film with the controlled electrochemical potential of mobile component, and the control is performed by the oxygen pressure. The linear equation of state $\boldsymbol{D} = \varepsilon_0 \varepsilon_d \boldsymbol{E}$ relates an electric displacement $\boldsymbol{D}$ and field $\boldsymbol{E}$ in the gap. Here $\varepsilon_0$ is a universal dielectric constant and $\varepsilon_d \sim (1 - 10)$ is a relative permittivity in the gap filled by an air with controllable oxygen pressure. A wide band-gap AFE film can be considered insulating, and hence within the material $\boldsymbol{D} = \varepsilon_0 \boldsymbol{E} + \boldsymbol{P}$. A potential $\phi$ of a quasi-static electric field inside the film satisfies a Laplace equation in the gap and a Poisson equation in the film, with the bound charge density determined by the polarization gradient. The boundary conditions for the system are the equivalence of the electric potential to the applied voltage $U$ at the top electrode (modeled by a flattened region $z = -\lambda$); and the equivalence of the difference $D_3^{(gap)} - D_3^{(film)}$ to the ionic surface charge density $\sigma[\phi(\vec{r})]$ at $z = 0$; and the continuity of the $\phi$ at gap - film interface $z = 0$; and zero potential at the conducting bottom electrode $z = h$.

The polarization components of the AFE film depend on the electric field $E_i$ as $P_i = P_i^f + \varepsilon_0 (\varepsilon_{ij}^b - 1) E_j$, where $i = 1,2,3$, and $\varepsilon_{33}^b$ is a relative background permittivity of antiferroelectric, $\varepsilon_{ij}^b \lesssim 10$[56]]. The polarization component $P_i^f$ is further abbreviated as $P_i$. To determine the spatial-temporal evolution of $P_i$, we use either the simplest two-sublattice Kittel model[57], or more complicated models incorporating antipolar modes related oxygen octahedra rotations[58–60] combined with the LGD approach. Corresponding LGD free energy functional $F$ additively includes a bulk part – an expansion on powers of 2-4 of the polarization ($P_i$) and anti-polarization ($A_i$), $F_{bulk}$; a polarization gradient energy contribution, $F_{grad}$; an electrostatic contribution, $F_{el}$; and a surface energy, $F_S$. It has the form:

$$F = F_{bulk} + F_{grad} + F_{el} + F_S, \tag{1}$$

where the constituent parts are

$$F_{bulk} = \int_{V_f} d^3 r \left[ \frac{a_i}{2} P_i^2 + \frac{a_{ij}}{2} P_i^2 P_j^2 + \frac{a_{ijk}}{4} P_i^2 P_j^2 P_k^2 + \frac{\chi_{ij}}{2} P_i^2 A_j^2 + \frac{b_i}{2} A_i^2 + \frac{b_{ij}}{2} A_i^2 A_j^2 + \frac{b_{ijk}}{4} A_i^2 A_j^2 A_k^2 \right], \tag{2a}$$

$$F_{grad} = \int_{V_f} d^3 r \frac{g_{ijkl}}{2} \left( \frac{\partial P_i}{\partial x_j} \frac{\partial P_k}{\partial x_l} + \frac{\partial A_i}{\partial x_j} \frac{\partial A_k}{\partial x_l} \right), \tag{2b}$$

$$F_{el} = - \int_{V_f} d^3 r \left( P_i E_i + \frac{\varepsilon_0 \varepsilon_b}{2} E_i E_i \right), \tag{2c}$$

$$F_S = \frac{1}{2} \int_S d^2 r \, a_{ij}^{(S)} \left( P_i P_j + A_i A_j \right). \tag{2d}$$

Here $V_f$ is the film volume. The coefficients $a_i$ and $b_i$ linearly depends on temperature $T$; in particular $a_i = \alpha_T (T_P - T)$ and $b_i = \alpha_T (T_A - T)$, so they change their sign at Curie temperature $T_P$ and AFE temperature $T_A$, respectively, $\alpha_T > 0$. All other tensors in Eqs.(2) are regarded as temperature-independent. The tensors $a_{ijk}$ and $b_{ijk}$, and the gradient coefficients tensor $g_{ijkl}$ are positively defined for the functional stability. The elastic and electrostriction energies are negligibly small for a single-domain film on a matched substrate. Otherwise, the influence of elastic strain can be approximately taken into account by the renormalization of coefficients in Eq.(2a) (see e.g.[1]).

The energy $F_{el}$ is dependent on the surface charge density $\sigma(\phi)$, since the electric field is ultimately determined by the charge distribution in the system. Several models have been introduced for the charge variation.[39,47] In principle, these models can incorporate specific DOS and electric potential at the interface, but only few are sensitive to the changes of interfacial potential in a self-consistent manner.

Below, we use an equation for the surface ionic charge density $\sigma(\phi)$ analogous to the Langmuir adsorption isotherm[61] and a linear relaxation model to describe the temporal dynamics of the surface charge density:

$$\tau \frac{\partial \sigma}{\partial t} + \sigma = \sigma_0[\phi]. \tag{3}$$

Here the dependence of an equilibrium charge density $\sigma_0[\phi]$ on the electric potential $\phi$ is controlled by the concentration of surface ions, $\theta_i(\phi)$, at the interface $z = 0$ in a self-consistent manner, as proposed by Stephenson and Highland[49,50]:

$$\sigma_0[\phi] = \sum_{i=1}^{2} \frac{eZ_i \theta_i(\phi)}{N_i} \equiv \sum_{i=1}^{2} \frac{eZ_i}{N_i} \left(1 + \rho^{1/n_i} \exp\left(\frac{\Delta G_i^{00} + eZ_i\phi}{k_B T}\right)\right)^{-1}, \tag{4}$$

where $e$ is an elementary charge, $Z_i$ is the ionization degree of the surface ions/electrons, $1/N_i$ are saturation densities of the surface ions. A subscript $i$ designates the summation on positive ($i = 1$) and negative ($i = 2$) charges, respectively; $\rho = \frac{p_{O2}^{00}}{p_{O2}}$ is the relative partial pressure of oxygen (or other ambient gas)[49,14], $n_i$ is the number of surface ions created per gas molecule. Two surface charge species exist, since the gas molecule had been electroneutral before its electrochemical decomposition started. The dimensionless ratio $\rho$ is chosen to vary in a range from $10^{-6}$ to $10^{6}$.

Positive parameters $\Delta G_1^{00}$ and $\Delta G_2^{00}$ are the free energies of the surface defects formation at normal conditions, $p_{O2}^{00} = 1$ bar, and zero applied voltage $U = 0$. The energies $\Delta G_i^{00}$ are responsible for the formation of different surface charge states (ions, vacancies, or their complexes). Specifically, exact values of $\Delta G_i^{00}$ are poorly known even for many practically important cases, and so hereinafter they are regarded varying in the range ~(0 − 1) eV[49]. Note that exact values of the parameters can be derived either via variable pressure experiments or by exploring polarization dynamics under electrochemical control.[62] However, we refer the theory-experiment matching to further studies.

Since the stabilization of a single-domain polarization in ultrathin perovskite films can take place due to the chemical switching (see e.g. Refs.[49,50,63]), we can assume the same for an AFE film. Also, we will assume that the distributions of $P_i(x, y, z)$ and $A_i(x, y, z)$ do not deviate significantly from their values averaged over the film thickness, which are further abbreviated as "polarization" $P_i \cong \langle P_i \rangle$ and "anti-polarization" $A_i \cong \langle A_i \rangle$. Note that this mean-field approximation reduces the problem to zero dimensional, and allows for analytical treatment as performed here.

In this case, the behavior of the polarization $P_i$, anti-polarization $A_i$, can be described via relaxation-type nonlinear differential equations,

$$\frac{\partial f}{\partial P_i} = -\Gamma_P \frac{\partial P_i}{\partial t}, \quad \frac{\partial f}{\partial A_i} = -\Gamma_A \frac{\partial A_i}{\partial t}. \tag{5}$$

Hereinafter we assume that the in-plane electric field components are absent, $E_1 = E_2 = 0$, and consider only the component $E_3$ in Eqs.(5). The expression for electric field $E_3$ follows from the minimization of electrostatic energy[55]:

$$E_3 = \frac{\Psi}{h} = \frac{\lambda(\sigma_0[\Psi] - P_3) + \varepsilon_0 \varepsilon_d U}{\varepsilon_0(\varepsilon_d h + \lambda \varepsilon_{33}^b)}. \tag{6}$$

Eq.(6) corresponds either to the stationary case, or to the adiabatic conditions, when $\sigma = \sigma_0[\Psi]$ in Eq.(3). The overpotential $\Psi$ contains the contribution from surface charges proportional to $\sigma_0$, the depolarization field contribution proportional to $P$, and the external potential drop proportional to $U$. Hence the component $E_3$ consists of the external field, depolarization field and the field created by the surface ion layer.

The constituent parts of the thermodynamic potential (1) can be further expanded in $A_i$, $P_i$ and $\Psi$ powers, assuming that $|eZ_i\Psi/k_BT| \ll 1$. In result we obtain that the surface ions and depolarization field change the coefficients $a_3$, $a_{i3}$, $\chi_{i3}$, induce the built-in field $E_{SI}$ and decrease the external field $E_a$ in the following way:

$$a_3(T,\rho,h) \to a_3\big(1 + S(T,\rho,h)\big) + \frac{\lambda}{2\varepsilon_0(\varepsilon_d h + \lambda\varepsilon_{33}^b)}, \tag{7a}$$

$$a_{i3}(T,\rho,h) \to a_{i3}\big(1 + S(T,\rho,h)\big), \quad \chi_{i3}(T,\rho,h) \to \chi_{i3}\big(1 + S(T,\rho,h)\big), \tag{7b}$$

$$E_3(T,\rho,h) = E_{SI}(T,\rho,h) + E_{act}(U,h), \tag{7c}$$

$$E_{SI}(T,\rho,h) = \frac{\lambda}{\varepsilon_0(\varepsilon_d h + \lambda\varepsilon_{33}^b)}\sum_{i=1,2}\frac{eZ_i}{N_i}f_i(T,\rho), \qquad E_{act}(U,h) = -\frac{\varepsilon_d U}{\varepsilon_d h + \lambda\varepsilon_{33}^b}. \tag{7d}$$

The last term in Eq.(7a), $\frac{\lambda}{2\varepsilon_0(\varepsilon_d h + \lambda\varepsilon_{33}^b)}$, originated from the depolarization field. Since, as a rule, $\varepsilon_d h \gg \lambda\varepsilon_{33}^b$, the acting field is close to an external field, $E_{act} \approx -\frac{U}{h}$. Also, we introduce positive functions in Eq.(7):

$$S(T,\rho,h) = \frac{\lambda h}{\varepsilon_0(\varepsilon_d h + \lambda\varepsilon_{33}^b)}\sum_{i=1,2}\frac{(eZ_i f_i(T,\rho))^2}{N_i k_B T}, \tag{7e}$$

$$f_i(T,\rho) = \left(1 + exp\left(\frac{\Delta G_i^{00}}{k_B T}\right) + \rho^{1/n_i}\right)^{-1}. \tag{7f}$$

It is seen from the from the expressions (6) - (7) that possible solutions of relaxation Eqs.(5), and, in particular, the features of static and dynamic polarization (P-E) hysteresis loops depend on a great number of parameters, some of which can vary in a relatively wide range. Specifically, besides the dynamic variables, such as applied pressure $U$, temperature $T$ and partial oxygen pressure $\rho$, one can vary the film thickness $h$ and the gap thickness $\lambda$, as well as the energies $\Delta G_i^{00}$ of the surface defects formation and their saturation densities $1/N_i$. Not the least, the coupling constants $\chi_{ij}$ are poorly known, as are the fitting parameters.

Our preliminary analysis of the phase diagrams calculated for a simplified 2-4-LGD expansion (without 6-th powers in Eq.(2a)) show that there are multiple possible phases in the system, such as paraelectric (PE) and multiple ferroelectric-like (FE), antiferroelectric and ferroionic phases nontrivially located in the space of variables $\{T,\rho,U\}$ ruled by the other parameters, namely external ones ($h$ and $\lambda$), interfacial and material constants ($\Delta G_i^{00}$, $1/N_i$ and $\chi_{ij}$), see the sketch below for illustration of the influence of the film thickness $h$ on the phase diagram. We expect multiple solutions routes, especially in a dynamic case, and the specific route critically depends on the point in the above multi-parameter space, via e.g., interfacial conditions (controlled by $\lambda$), finite size effects (controlled by $h$), and, maybe the most intriguing via the surface ions coupling to the soft modes in the AFE film (controlled by parameters $\Delta G_i^{00}$, $1/N_i$ and $\chi_{ij}$). These control parameters define the multidimensional parameter space of the system.

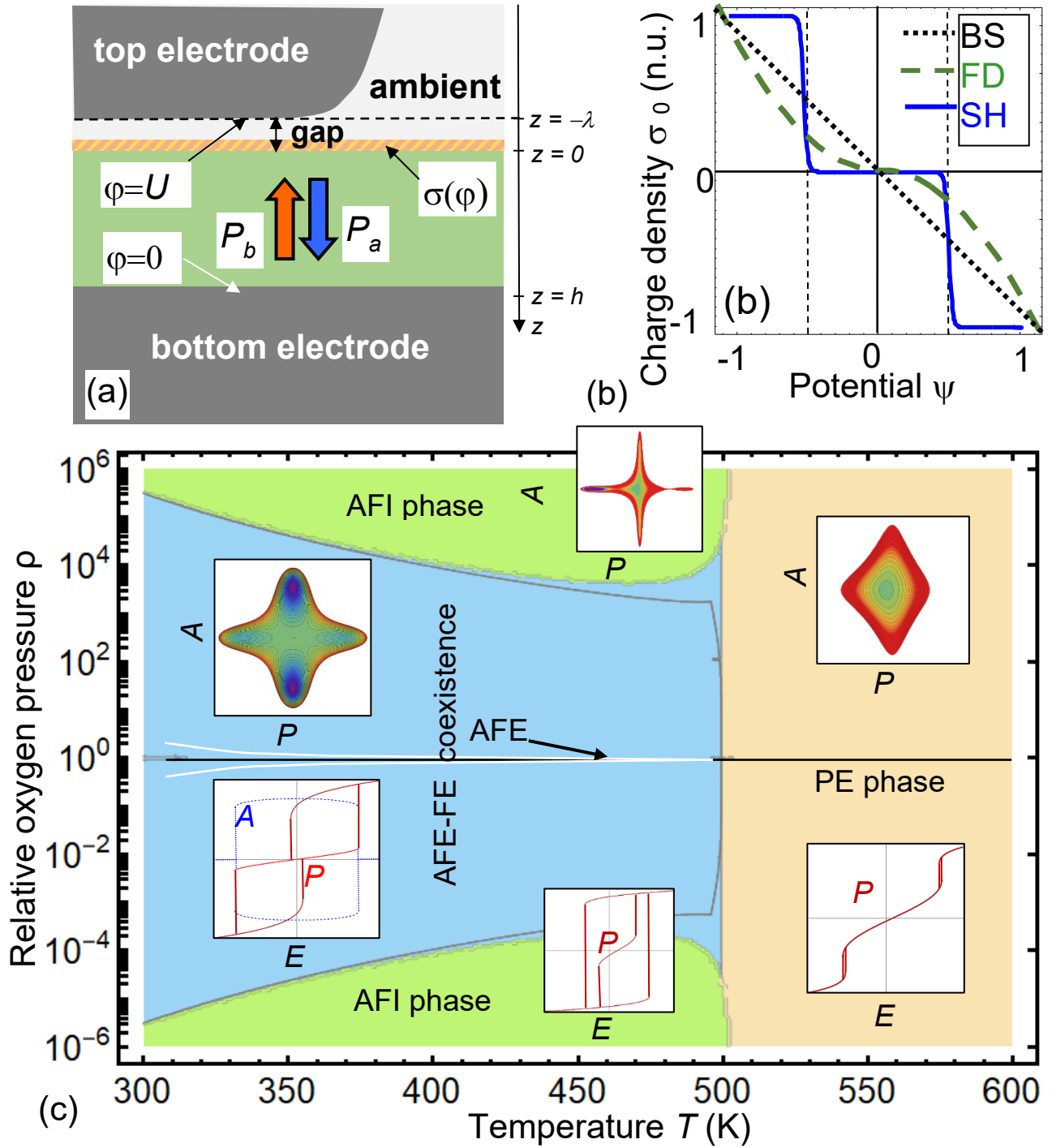

**Figure 1:** **(a)** Layout of the considered system, consisting of electron conducting bottom electrode, ferroelectric (AFE) film, screening layer with charge density $\sigma(\varphi)$, and ultra-thin gap separating the film surface and the top electrode. **(b)** Schematic dependence of the normalized charge densities $\sigma_0$ vs. dimensionless electric potential $\psi$ calculated for the models of Bardeen-type surface states (black line "BS") and Fermi-Dirac density of states (red curve "FD") describing 2D electron gas, in comparison with the Stephenson-Highland model describing the surface charge density of absorbed ions (blue step-like curve "SH"). Adapted from [47] **(c)** A typical phase diagram of an AFE film in dependence on the temperature $T$

and relative partial oxygen pressure $\rho$. There are an antiferroelectric (AFE) phase coexisting with a weak ferroelectric (FE) phase, a ferroelectric-like antiferroionic (AFI) phase, and an electret-like paraelectric (PE) phase. Free energy maps at zero electric field $E = 0$ (upper insets) and the polar ($P$) and antipolar ($A$) order parameters dependence on the static $E$-field (bottom insets), are calculated in each phase. Adapted from [55].

Practically of interest for applications are the macroscopic responses of the system, including energy storage and losses for a given film thickness and bias window. With the formalism above, the relevant P-E loops can be readily evaluated, and parameters can be extracted. However, the relationships between these can be highly non-trivial, with a change in a single parameter affecting multiple aspects of materials functionality. Combined with the high-dimensional nature of the parameter space, this precludes the simple gradient descent or grid search methods. This necessitates the introduction of Bayesian Optimization methods capable to combine exploration and exploitation of the relevant parameter spaces. Previously, we adopted these approaches for single parameter optimization of ferroelectric materials[48] and lattice models[1] and theory experiment matching.[64] Below, we discuss the basic principles of BO and especially the Multi Objective BO, necessary for discovery balancing multiple functionalities.

## 2. Bayesian optimization (BO) and Multi-objective Bayesian optimization (MOBO)

Bayesian optimization[65] (BO) is an emerging field of study in the Sequential Design Methods. It has been considered as a low computationally cost global optimization tool for design problems having expensive black-box objective functions. The general idea of BO is to emulate unknown functional behavior in the given parameter space and find the local and global optimal locations, while reducing the cost of function evaluations from expensive high-fidelity models. The reason it is called Bayesian is that it follows the ideology of Bayes theorem, which states that "posterior probability of a model (or parameters) M given evidence (or data, or observations) E is proportional to the likelihood of E given M, multiplied by the prior probability of E." Mathematically:

$$p(M|E) \propto \ell(E|M)p(M) \tag{8}$$

In the BO setting, the prior represents the belief of the unknown function $f$, assuming there exists a prior knowledge about the function (e.g. smoothness, noisy or noise-free etc.). Given the prior belief, the likelihood represents how likely is the observed data, $D$. These data are the sampled data as the true function value from expensive evaluations and can be viewed as the realizations from the unknown function. Finally, given the data, the posterior distribution is computed in the BO as a posterior surrogate model (e.g. Gaussian process model (GPM)): $\Delta = p(f|(D))$ is developed from these sampled data. Thus, Eqn. 9 in the Bayesian optimization setting, can be modified as below:

$$p(f|(D_{1:k})) \propto \ell((D_{1:k})|f)p(f) \tag{9}$$

where $D_{1:k} = [x_{1:k}, f(x_{1:k})]$ is the augmentation of the observation or sampled data till $k^{th}$ iteration of BO. Unlike the general Bayesian modelling, here prior or likelihood function is not written separately, but the augmentation of data at each iteration combines the prior distribution with the likelihood function. Generally, the GPM posterior model is considered based on conjugate normal prior assuming the data is realized from the Gaussian (normal) function.

This approach has been widely used in many machine learning problems[66–70]. However, attempts have been made where the response is discrete such as in consumer modeling problems where the responses are in terms of user preference.[65,71] The idea is to approximate the user preference discrete response function into continuous latent functions using Binomial-Probit model for binary choices[72,73] and polychotomous regression model for more than two choices where the user can state no preference.[74] In the domain of material science, though the optimization problem in material characterization and syntheses have expensive (computational) or unknown (physical experiments) functions, the application of BO has not progressed much yet[75]. In some notable collaboration between BO and material science, the researchers found the similar performance with lot lesser BO evaluations than exhaustive evaluations in optimizing different properties of materials like grain boundary structure, grain boundary energy, elastic properties etc., thus reducing cost and increasing efficiency.[76–79] Even in the functions are computationally cheap, BO allows for quick convergence for very large parameter space than doing exhaustive search.[80] Furthermore, BO is also applied in optimizing hysteresis loop shape in ferroelectric materials using continuous lattice model.[81] Just like in expensive computational simulations, BO also helped in reducing the effort and resource cost in experimental material science problems.[82–84] However, the vast majority of these applications address only single-parameter optimization, whereas in many cases we aim to discover the optimal balance between two or more functionalities, as implemented in Multi Objective BO. To date, there are only few applications of MOBO for material discovery in optimizing multiple target properties, considering both experimental and computer simulated data[85]. MOBO has also been applied for efficient discovery of the targeted materials, performed by a thermodynamically-consistent micromechanical model that predicts the materials response based on its composition and microstructure.[50]

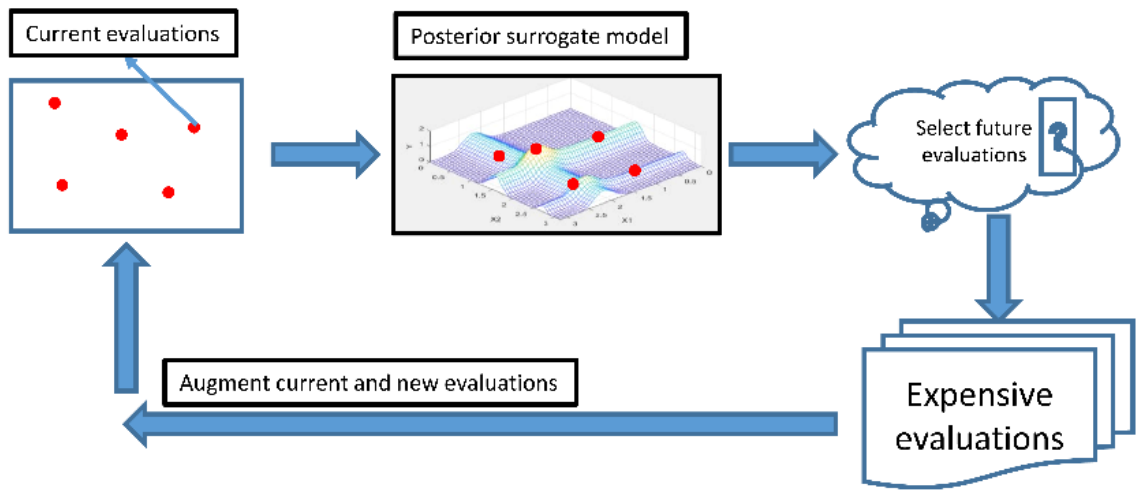

**Figure 2.** Bayesian Optimization Framework.

BO adopts a Bayesian perspective and assumes that there is a prior on the function; typically, a Gaussian process. The prior is represented from the true model or experiments which is assumed as the realizations of the true function, is treated as training data. The overall Bayesian Optimization Approach has two major components: A predictor or Gaussian Process Model (GPM) and an Acquisition Function (AF). As shown in figure 2, in this approach, a posterior GPM is first built or iteratively updated, given the data from the current evaluations (training data). The surrogate GPM then predicts the realizations of the function at the unexplored locations over the control parameter space. The best locations are then strategically selected in the space for future expensive evaluations by maximizing the acquisition functions, defined from the posterior GPM simulations. Finally, after the evaluations are done, they are augmented with the existing evaluations and the cycle is repeated till the model is converged.

It is important to mention that as an alternative to a GPM, random forest regression has been proposed as an expressive and flexible surrogate model in the context of sequential model-based algorithm configuration.[51] Although random forests are good interpolators in the sense that they output good predictions in the neighborhood of training data, they are very poor extrapolators where the training data are far away.[52] This can lead to selecting redundant exploration (more experiments) in the non-interesting region as suggested by the acquisition function in the early iterations of the optimization, due to having additional prediction error of the region far away from the training data. This motivates us to consider the GPM in a Bayesian framework while extending the application to complex physics driven problems.

The numerical optimization problems can be classified on the basis of number of objective functions as Single (SOO) and Multi-Objective Optimization problems (MOO). MOO is the extension of SOO with having more than one objective as eqn. 10.

$$\max f(X) = [\max f_1(X), \max f_2(X), \dots \max f_n(X)] \ s.t \ X \in \mathbb{R} \tag{10}$$

It is obvious that SOO is relatively simpler with lower computational cost; however, in practical problems it may be a challenge to formulate a single objective problem and therefore, much focus has been given on Multi-Objective Optimization methods (MOO). The question generally attempted to solve in any MOO is the optimal decisions under user defined preferences of objectives, which are referred as pareto-optimal solutions. Considering a minimization (maximization) problem, a candidate point is a Pareto-optimal solution if there is no other feasible point which gives a lower (higher) minimum (maximum) objective function for at least one of the objectives without the sacrifice from others. Such pareto-optimal solutions at different trade-offs of objectives are represented by a Pareto frontier. Pareto-optimal solutions are also referred as non-dominated solutions in the design or objective criterion space. The methods to solve MOO problems can be classified into a priori and a posteriori Methods based on whether we predefine the weightage associated to each objective or not respectively. Similarly, in the Bayesian framework, MOBO is the extension of BO, where we focus on optimizing multiple expensive responses or objective functions and build a Pareto frontier. We next describe the GPM and AF.

2.1. *Gaussian process model (GPM)*

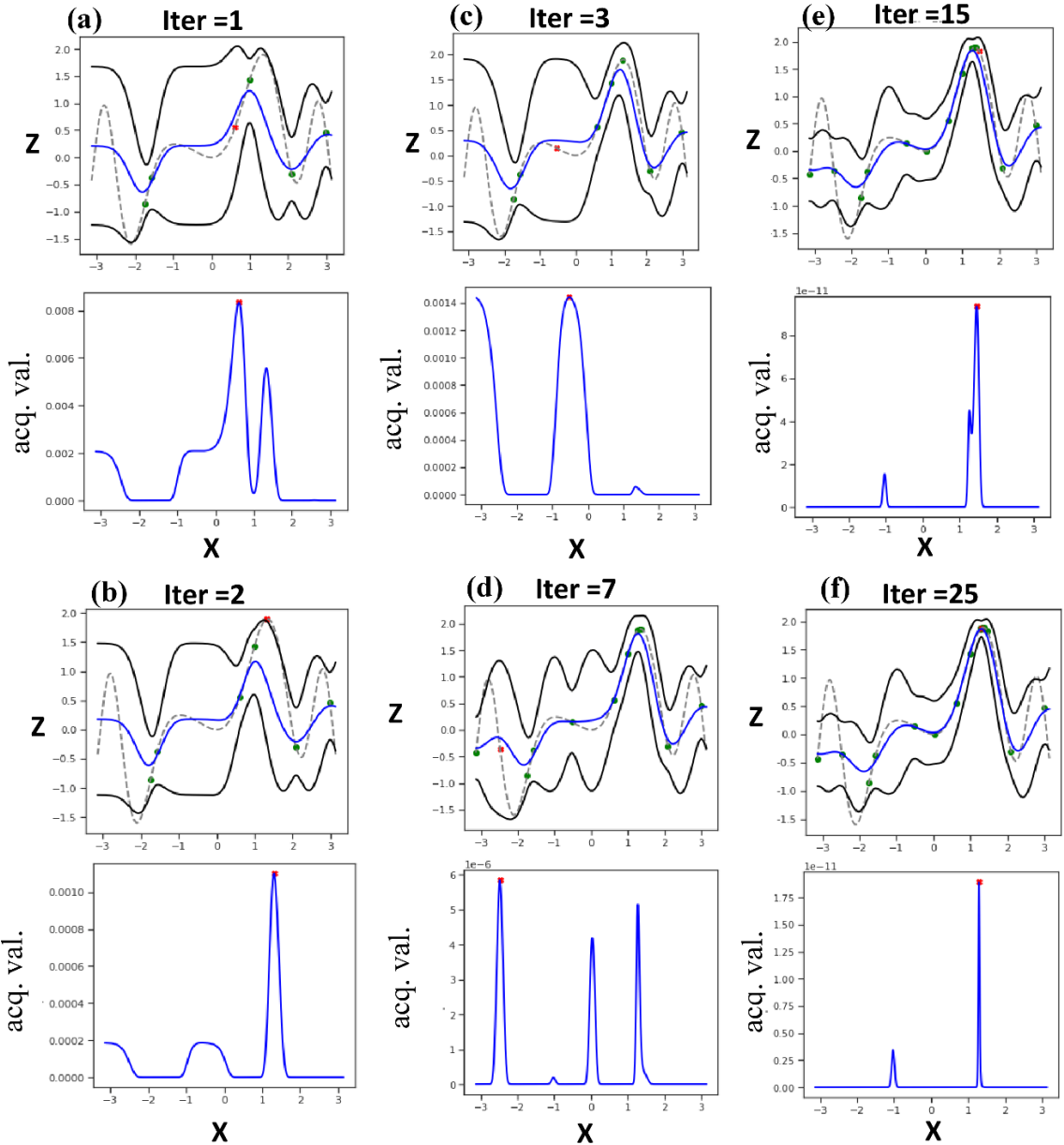

**Figure 3.** 1D Gaussian Process and search space exploration by maximizing acquisition function. The top images of (a-f) are the Gaussian process illustration. Gray dotted lines are the true function value, blue solid lines are the GP predicted mean, black solid lines are the GP mean ± 2x standard dev. Green dots are the current evaluated locations. Red dots are the new locations for next evaluations. The bottom images of (a-f) are the acquisition function illustration. Blue solid lines are acquisition function. Red dots are the new locations (at max. acq. func. value) for next evaluations.

As a simple example, figure 3 (a-f) shows a simple 1D Gaussian Process Model with one control parameter $x$ and one objective function variable $z = f(x)$. The green dots are the evaluated locations, and the grey dotted and blue solid lines are the true and the predictor mean functions in the parameter space, given the prior evaluated data. The area enclosed by the black curves shows the measure of uncertainty over the surrogate GPM prediction. It is clearly seen that the variance near the observations is small and increases as the design samples are farther away from the evaluated data, thereby related to kriging models where the errors are not independent. Much research has been ongoing regarding incorporating and quantifying uncertainty of the experimental or training data by using a nugget term in the predictor GPM. It has been found that the nugget provides better solution and computational stability framework.[89,90] Furthermore, GPM has also been implemented in high dimensional design space exploration[91] and BIG DATA problems[92], as an attempt to increase computational efficiency. A survey of implementation of different GP packages has been provided in different coding languages such as MATLAB, R, Python[93]. Extending BO to MOBO, we have multiple Gaussian priors, one for each expensive functions, and therefore we develop respective multiple posterior GPMs. Since, in classical MOO problems, each functions is treated independently during optimization, here also in MOBO, we consider multiple independently fitted GPMs.

The general form of the GPM is as follows:

$$y(x) = x^T\beta + z(x) \tag{11}$$

where $x^T\beta$ is the Polynomial Regression model. The polynomial regression model captures the global trend of the data. In general, 1st order polynomial regression is used, which is also known as universal kriging[94]; however, it has also been claimed that it is fine to use a constant mean model.[95] $z(x)$ is a realization of a correlated Gaussian Process with mean $E[z(x)]$ and covariance $cov(x^i, x^j)$ functions defined as follows:

$$z(x) \sim GP\left(E[z(x)], cov(x^i, x^j)\right); \tag{12a}$$

$$E[z(x)] = 0, cov(x^i, x^j) = \sigma^2 R(x^i, x^j) \tag{12b}$$

$$R(x^i, x^j) = \exp\left(-0.5 * \sum_{m=1}^{d} \frac{\left(x_m^i - x_m^j\right)^2}{\theta_m^2}\right); \tag{12c}$$

$$\theta_m = (\theta_1, \theta_2, \dots, \theta_d)$$

where $\sigma^2$ is the overall variance parameter and $\theta_m$ is the correlation length scale parameter in dimension $m$ of $d$ dimension of $x$. These are termed as the hyper-parameters of GP model. $R(x^i, x^j)$ is the spatial correlation function. In this paper, we have considered the constant mean model and a Radial Basis function which is given by eqn. 12c. The objective is to estimate (by MLE) the hyper-parameters $\sigma$, $\theta_m$ which creates the surrogate model that best explains the training data $\boldsymbol{D_k}$ at iteration $k$.

After the GP model is fitted, the next task of the GP model is to predict at an arbitrary (unexplored) location drawn from the parameter space. Assume $\boldsymbol{D_k} = \{\boldsymbol{X_k}, \boldsymbol{Y(X_k)}\}$ is the prior information from previous evaluations or experiments from high fidelity models, and $\bar{\bar{x}}_{k+1} \in \bar{\bar{\boldsymbol{X}}}$ is a new design within the unexplored locations in the parameter space, $\bar{\bar{\boldsymbol{X}}}$. The predictive output distribution of $x_{k+1}$, given the posterior GP model, is given by eqn. 13a.

$$P(\bar{y}_{k+1}|\boldsymbol{D_k}, \bar{\bar{x}}_{k+1}, \sigma_k^2, \boldsymbol{\theta_k}) = \boldsymbol{N}(\mu(\bar{y}_{k+1}(\bar{\bar{x}}_{k+1})), \sigma^2(\bar{y}_{k+1}(\bar{\bar{x}}_{k+1}))) \tag{13a}$$

where:

$$\mu\big(\bar{\bar{y}}_{k+1}(\bar{\bar{x}}_{k+1})\big) = cov_{k+1}^T COV_k^{-1} Y_k; \tag{13b}$$

$$\sigma^2\big(\bar{\bar{y}}_{k+1}(\bar{\bar{x}}_{k+1})\big) = cov(\bar{\bar{x}}_{k+1}, \bar{\bar{x}}_{k+1}) - cov_{k+1}^T COV_k^{-1} cov_{k+1} \tag{13c}$$

$COV_k$ is the kernel matrix of already sampled designs $X_k$ and $cov_{k+1}$ is the covariance function of new design $\bar{\bar{x}}_{k+1}$ which is defined as follows:

$$COV_k = \begin{bmatrix} cov(x_1, x_1) & \cdots & cov(x_1, x_k) \\ \vdots & \ddots & \vdots \\ cov(x_k, x_1) & \cdots & cov(x_k, x_k) \end{bmatrix}$$

$$cov_{k+1} = [cov(\bar{\bar{x}}_{k+1}, x_1), cov(\bar{\bar{x}}_{k+1}, x_2), .., cov(\bar{\bar{x}}_{k+1}, x_k)]$$

## 2.2. *Acquisition function (AF)*

The second major component in Bayesian optimization is the Acquisition Function (AF). AF guides the search for future evaluations or experiments towards the desired goal and thereby bring the sequential design into the BO. The AF predicts an improvement metric for each sample. The improvement metric depends on exploration (probability to discover useful behaviors in unexplored locations) and exploitation (known region with high responses). Thus, the acquisition function gives high value of improvement to the samples whose mean prediction is high, variance is high, or a combination of both. By maximizing the acquisition function, we select the best samples to find the optimum solution and reduce the uncertainty of the unknown expensive design space. Figure 3 (a-f) shows a simple example of BO exploration with one control parameter x and one objective function variable $z = f(x)$ of the sequential selection of samples by maximizing the acquisition function, given posterior GP model in iterations 1-3, 7, 15 and 25. The solid blue line is the GP predicted function and the dotted grey line is the original function. The green dots in iteration 1 (3a. top figure) are the evaluated locations and the red dot is the new evaluated location, as per maximizing the acquisition function (3a. bottom figure). The new data is augmented with current data and the GPM is updated in the next iteration, and then the acquisition function is maximized to get the next evaluated locations (see fig. 3b). This process is repeated iteratively, and in this manner the search space is explored. We can see from the figure that the acquisition function value is highest where the samples have high prediction mean and/or high variance and the lowest where the samples have low prediction, low variance or both. Thus, the acquisition function strategically selects points which have the potential to be optimal (eg. maximum value of the unknown function) and gradually reduces the error and aligns with the true function, at those potential region, with the iterations.

Throughout the years, various formulations have been applied to define the acquisition functions. One such method is the Probability of Improvement, PI[96] which is improvement-based acquisition function. Jones[97] notes that the performance of PI($\cdot$) "is truly impressive;… however, the difficulty is that the PI($\cdot$) method is extremely sensitive to the choice of the target. If the desired improvement is too small, the search will be highly local and will only move on to search globally after searching nearly exhaustively around the current best point. On the other hand, if the small-valued tolerance parameter ξ in PI(.) equation is set too high, the search will be excessively global, and the algorithm will be slow to fine-tune any promising solutions." Thus, the Expected Improvement acquisition function, EI[65], is widely used over PI which is a trade-off between exploration and exploitation. Another Acquisition function is the Confidence bound criteria, CB, introduced by Cox and John[98], where the selection of points is based on the upper or lower confidence bound of the predicted design surface for maximization or minimization problem respectively. However, these acquisition functions are limited for the case when there is a single optimization functionality. Hence, we further discuss the acquisition functions for MOBO, where the target is to find the manifold in the parameter space of the system containing the points on the Pareto frontier for the system.

In MOBO, the computation of the acquisition function is more complicated and can be computationally costly. Currently, there are two approaches to configure the acquisition functions in MOBO, namely priori method and posteriori method. The two differ in whether the weighting factors for each objective are predefined or not. This is similar technique like classical MOO problems. However, unlike MOO where the weighting factors are associated to objective functions, in MOBO, the same are associated to acquisition functions. In the posteriori approach, the goal is to maximize hyper-volume indicator function to maximize the likelihood of iteratively selecting non-dominated solutions (green dots) as shown in fig. 4. The shaded area is the hyper-volume at iteration 1. As we proceed from iteration 1 to 2, we found three additional non-dominated solutions, where two of the earlier non-dominated solutions are now dominated (blue dots), which guaranteed that those solutions are not Pareto optimal. Eventually, we see increase of hypervolume in iteration 2 (red colored area). This reselection of dominated solutions goes on iteratively until the convergence criteria is met, which finally configures the pareto-frontier by the set of current dominated solutions. Different methods in this approach are Expected Improvement Hyper-volume (EIHV), Expected Hyper volume gradient-based (EIHVG), Max-value Entropy Search (MESMO) and Predictive Entropy Search (PESMO) acquisition functions. EIHV have been modeled to provide better performance[99,100]. However, to increase the computational efficiency, Yang et al.[101] modified the EIHV acquisition function into the EIHVG acquisition function and proposed an efficient algorithm to calculate it. To reduce the computational cost, MESMO[102] and PESMO[103] have been formulated. Abdolshah et. al.[104] proposed a multi-objective BO framework with preference over objectives on the basis of stability and attempting to focus on the Pareto front where preferred objectives are more stable. However, due to the computational complexity of all these MO-BO approaches with increasing number of objectives, the weighted Tchebycheff method has been augmented in MOBO with a ridge regularization term to smoothen the converted single multi-objective function[105]. Likewise, a weighted Tchebycheff method has been implemented in MO-BO and introduced a regression-based model calibration to estimate the parameter (utopia) of the multi-objective function[106]. Here, the weights are predefined by the user. The architecture is further modified with optimizing the regression-based model for calibration from the ensemble of models in order to enhance the model performance[107]. Another priori method is ParEGO[108], where the weights are randomly assigned with a uniform distribution. In these priori methods[106–108], the multiple objectives are first transformed into a single weighted multi-objective function and then fit into a single-function based acquisition function (for example – EI).

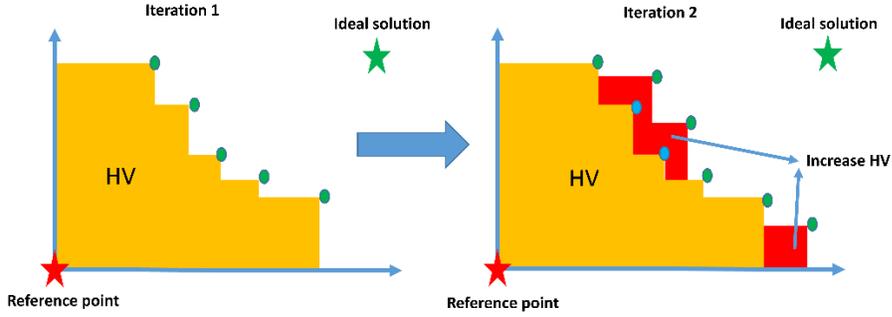

**Figure 4**. Hyper-volume computation in MOBO

### 2.4. *Multi-objective Bayesian optimization framework (MOBO)*

Table 1 presents the high level description of MOBO algorithm. For the sake of simplicity, we considered 2 control parameters (2D) and 2 objective functions. However, the model can be easily extended to 'n' parameters and 'j' objective functions.

*Table 1: Algorithm: Multi-objective Bayesian optimization (MOBO)- 2D and 2 objectives*

1. **Initialization:** Define parameter space. State maximum iteration, $N$. Randomly select $m$ samples, $\boldsymbol{X} = \{\boldsymbol{X_1}, \boldsymbol{X_2}\}$. Assuming $f_1, f_2$ are the two expensive objective functions. Evaluate $m$ samples for both objectives as, $\boldsymbol{Y_1}(\boldsymbol{X_k})$ and $\boldsymbol{Y_2}(\boldsymbol{X_k})$ respectively. Build training data matrices, $\boldsymbol{D_{1,k}} = \{\boldsymbol{X_k}, \boldsymbol{Y_{1,k}}\}$ and $\boldsymbol{D_{2,k}} = \{\boldsymbol{X_k}, \boldsymbol{Y_{2,k}}\}$. Set $k = 1$.

For $k \leq N$

2. **Surrogate Modelling**: Develop or update GPM models for both objectives as $\Delta_1(\boldsymbol{D_{1,k}})$ and $\Delta_2(\boldsymbol{D_{2,k}})$.
    a. Optimize the hyper-parameters of GPM by minimizing the loss (negative marginal log-likelihood) function using Adam optimizer algorithm. Here, we consider learning rate 0.2.

3. **Posterior Predictions**: Given the surrogate model, compute posterior means and variances for the unexplored locations, $\overline{\overline{X_k}}$, over the parameter space as $\{\boldsymbol{\pi_1}(Y(\overline{\overline{X_k}})|\Delta_1, \boldsymbol{\pi_2}(Y(\overline{\overline{X_k}})|\Delta_2\}$ and $\{\boldsymbol{\sigma_1^2}(Y(\overline{\overline{X_k}})|\Delta_1, \boldsymbol{\sigma_2^2}(Y(\overline{\overline{X_k}})|\Delta_2\}$ respectively.

4. **Acquisition function:** Compute and maximize acquisition function, $\max_x U(f_1, f_2|\Delta_1, \Delta_2)$ to select next best location, $\boldsymbol{x_{best,k}}$ for evaluations.

5. **Augmentation:** Evaluate $y_1(\boldsymbol{x_{best,k}})$ and $y_2(\boldsymbol{x_{best,k}})$ and augment data, $\boldsymbol{D_{1,k+1}} = [\boldsymbol{D_{1,k}}; \{\boldsymbol{x_{best,k}}, \ y_1\}$ and $\boldsymbol{D_{2,k+1}} = [\boldsymbol{D_{2,k}}; \{\boldsymbol{x_{best,k}}, \ y_2\}$.

Here, we considered two acquisition functions,[107] namely: 1. qEIHV and 2. qParEGO. This is the extension of EIHV and ParEGO respectively where parallelism is introduced and we can jointly optimize to provide the next best location, $\boldsymbol{x_{best,k}}$ in batch such that $\boldsymbol{X_{best,k}} =$ $\{x_{best,k,1}, ..., x_{best,k,q}\}$. This reduces the computational effort and can be extremely helpful when it is desirable (lesser time) for expensive evaluations in batch. For this study, we choose batch size, $q = 4$. Obviously, the choice of acquisition function can change the performance of MOBO depending on the problem, however, choosing the best acquisition function is a hard problem and is not the scope of this paper and can be addressed in future. It is up to the knowledge of the user to select or build an appropriate acquisition function for a given problem.

To initialize the computation of the acquisition function, we select 20 restart initial locations for a set of 1000 random points, and considered maximum iterations of 200. The acquisition function is approximated using 500 Monte-Carlo (MC) based samples. To compute the hyper-volume indicator function in qEIHV, a reference point (red star in fig. 4), which is the lower bound of the objectives (worst solution since MOBO is set up as maximization problem), is required. We set this value (after normalizing the objectives) as $0 - \delta$. $\delta = 0.001$ is a small positive value to make the reference point slightly worse than true lower bound of the objectives. Different strategies can be applied to select reference point based on the problem complexity, especially when we don't have much knowledge on the lower bound of the objective functions (for constrained problems), to increase the computational efficiency of the acquisition function. However, this is beyond the scope for this paper and we have considered the stated reference point throughout our analysis since we have good knowledge on the lower bound of the objectives (normalized).

To illustrate the MOBO framework in Table 1, we implemented it on two numerical test problems. To see the general application of the proposed design architecture, we first implement using two numerical multi-objective examples. As the scope of the proposed Bayesian framework in this paper is for solving expensive multi-objective problems, we assume the numerical functions are expensive in nature. We have considered here two numerical test problems(maximization): Test Problem (1): *ZDT1* and Test Problem (2): *2-D Six-hump camel back - Inversed-Ackley's Path (6HC-IAP)* multi-objective functions. The full mathematical equations of the optimization problems are provided in Appendix. We start with 10 randomly selected samples, and set the max iteration of MOBO as 50. Thus, the total evaluation is 10+50x4 = 210. Though, we solved the problems, each with two acquisition functions, we present here the results (figs. 5-6) of MOBO for the acquisition function performed better.

Figures 5 (a, b), 6 (a, b) shows the ground truth of the individual functions ($f_1$, $f_2$) of test problem (1) and (2) respectively. The ground truth is configured with exhaustive grid based sampling in the parameter space where true evaluations is done at all the grid points on a 100x100 grid. Comparing individual functions (figs. a, b) of each of the test problems in figs. 5 and 6, we can see clear trade-offs between them (see the color representations - yellow region being the maximum value region of individual functions), which gives us Pareto frontier (optimal balance) as we attempt to optimize (eg. maximize) both functions jointly (optimizing trade-offs) by implementing MOBO.

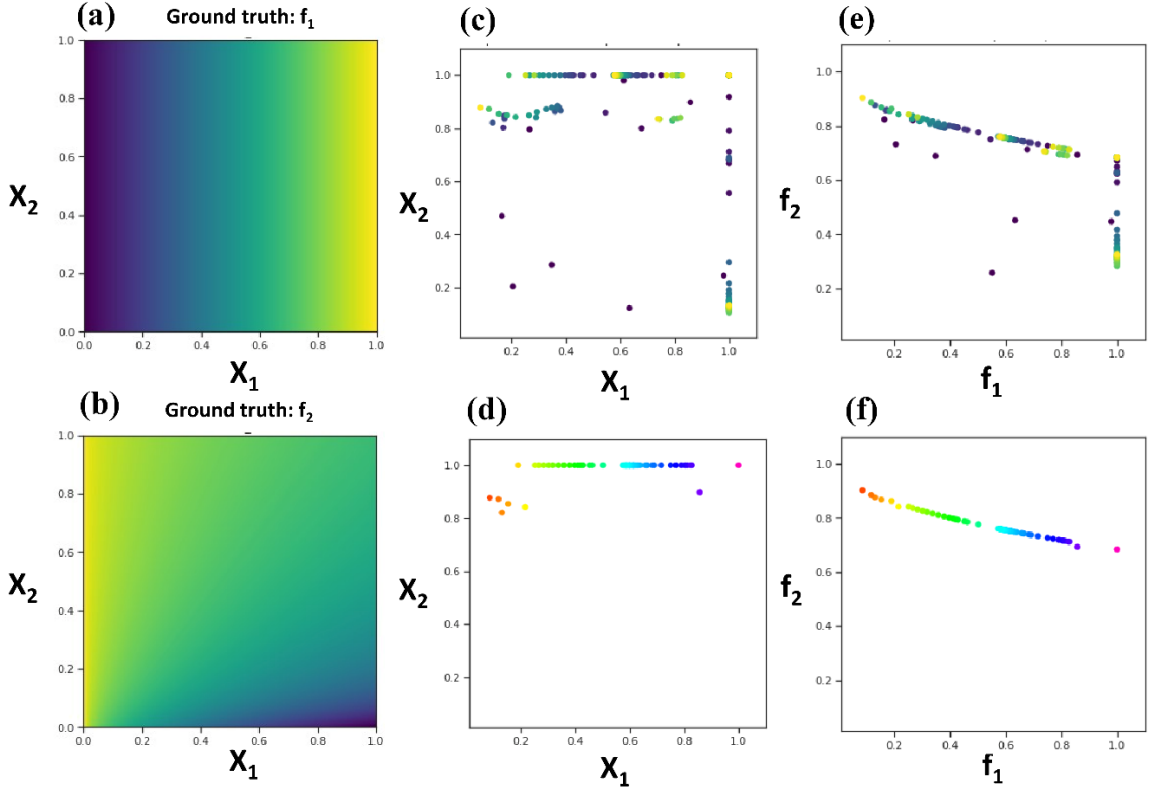

**Figure 5:** Test Problem: ZDT1. (a) Ground truth for function 1 (appendix eqs. 14). (b) Ground truth for function 2 (appendix eq. A.2). (c) All Samples selected after 50 MOBO iterations in real valued parameter space. (d) Pareto optimal solutions over real valued parameter space. (e) All Samples selected after 50 MOBO iterations in real valued objective space. (f) Pareto frontiers drawn over real valued objective space. Here, Acquisition function chosen: qEIHV.

Comparing the ground truth of the functions of both test problems, it can be easily interpreted that the test problem (2) is highly complex than test problem (1). Figures 5 (c, e), 6 (c, e) show all the samples selected through MOBO iterations in the parameter and objective space respectively for each of the test problem (1) and (2), using qEIHV and qParEGO acquisition functions respectively. The color coding is defined as the darker color resembles early iterations and the lighter color resembles later stage of iterations. Here, looking figs. 5e and 6e, we see most evaluations far away from the Pareto-frontier during the early stage of MOBO (explore), but then once it quickly identifies the region of interest (Pareto-frontier), exploits that region with the guidance of more evaluations near that region. This is power of adaptive sampling in MOBO, which maximizes the learning focused on region of interest through exploration-exploitation. Readers can look into figs. 5c and 6c to get an understanding of where most of the sampling has been done in the parameter space of ground truth. Also, we have seen that using the stated acquisition functions for the test problems, qParEGO does more exploration than qEIHV and therefore, has few more selection of samples far away from the pareto-frontier. The Pareto-optimal solutions, extracted from respective figs. 5c, 6c in the parameter space and build the pareto frontier in the objective space for both test problems, are drawn over the parameter space as shown in figs. 5d, 6d respectively. Figures 5f, 6f show the Pareto-frontiers in the objective space, extracted from the respective figures 5e, 6e. From the Pareto-frontier, one can visualize the trade-off between the objectives and can choose one or multiple solutions based on the desired trade-off.

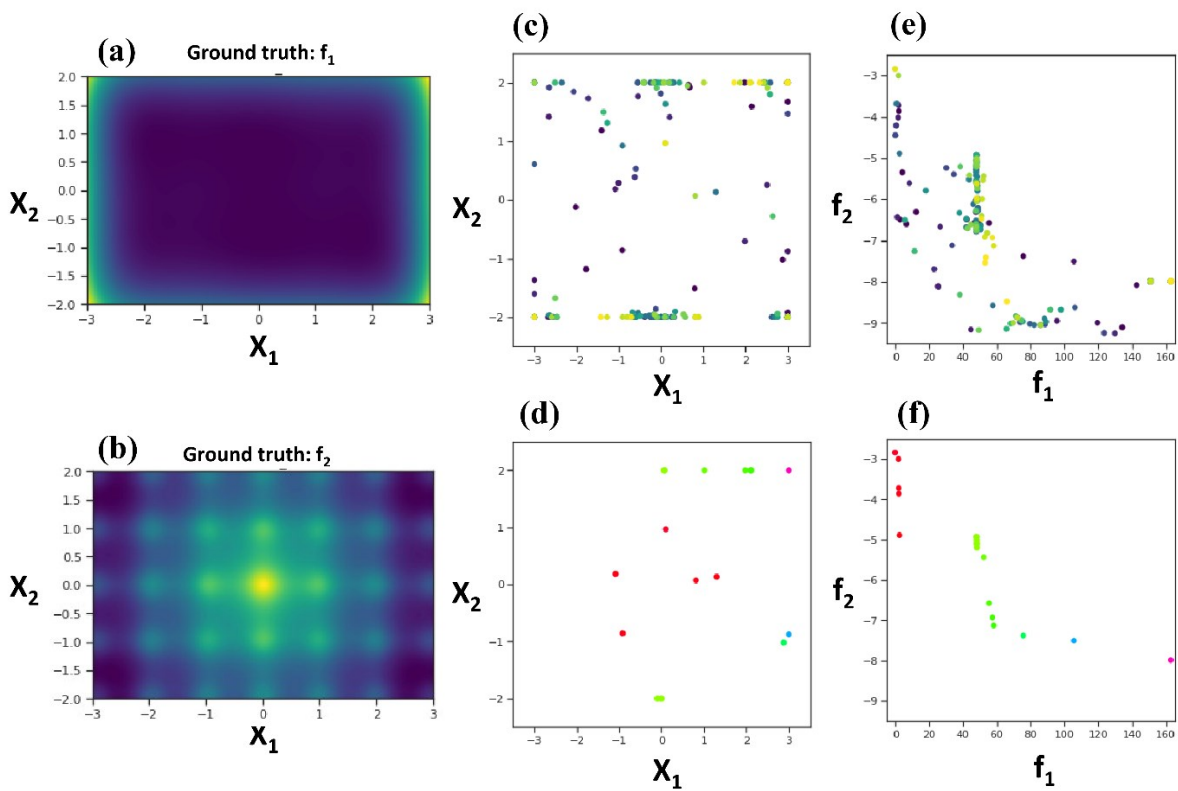

**Figure 6:** : Test Problem: 6HC-IAP. (a) Ground truth for function 1 (appendix eqs. 15). (b) Ground truth for function 2 (appendix eq. A.4). (c) All Samples selected after 50 MOBO iterations in real valued parameter space. (d) Pareto optimal solutions over real valued parameter space. (e) All Samples selected after 50 MOBO iterations in real valued objective space. (f) Pareto frontiers drawn over real valued objective space. Here, Acquisition function chosen: qParEGO.

To measure the performance of pareto-frontier of the test problems with MOBO, we compare with the respective pareto-frontier by solving Non-dominated sorting genetic algorithm II (NSGA-II) in '*pymoo*' package[110] in Python. We select the population size =100, and generation =100, total of 10000 function evaluations to ensure the configuration of the pareto-frontier. We consider this as the ground truth of the pareto frontier. Using MOBO, our goal is to cover this pareto frontier with maximizing accuracy and minimizing cost (lesser evaluations). Also, we run NSGA-II with only population size =15, and generation =15, total of 225 function evaluations, which is close to the number of function evaluations we did in MOBO. Figure 7 shows the comparison. Given that we did only 2% evaluation in MOBO that what we did in full convergence of NSGA-2 (210 function evaluations in MOBO, compared to 10000 evaluations in NSGA2), MOBO offered a superior performance in identifying a large portion of pareto-optimal solutions which overlaps with the ground truth in test problem ZDT1. Similarly, MOBO, with both acquisition functions, clearly outperforms by a long margin when the pareto-frontier is built with similar function evaluations in NGSA-2 (figs. 7 a, b). Obviously, 6HC-IAP problem is much more complex than ZDT1 as we can see from the ground truth of individual functions (figs. 6 a, b) and highly disjoint pareto frontier (10000 NGSA-2 evaluations), MOBO need more evaluations to cover more pareto-frontier region. However, even with only 210 evaluations for a complex problem with highly dis-jointed pareto-frontier, MOBO could identify pareto-solutions in 3 out of 4 pareto sub-frontiers (for qParEGO) and 4 out of 4 pareto sub-frontiers (for qEIHV), and did not stuck into one region. However, the pareto-optimal solutions are more accurate for qParEGO (figs. 7 c, d). Thus, the user though won't have full knowledge of the pareto-frontier but will have a good approximated idea over the whole objective or parameter space. Here also, MOBO specially with qParEGO acquisition function found more pareto optimal solutions than running similar evaluations with NGSA-2. Obviously, in case of expensive functions, it won't be possible for 10000 expensive evaluations, thus we are confident enough with the performance of MOBO in terms of the trade-off between accuracy and cost. More detailed figures on the test problems, with the iteration (steps-wise) of MOBO is provided in Colab notebook. Next, we proceed to solve a complex physics driven model using MOBO for performance optimization for memory and energy storage applications.

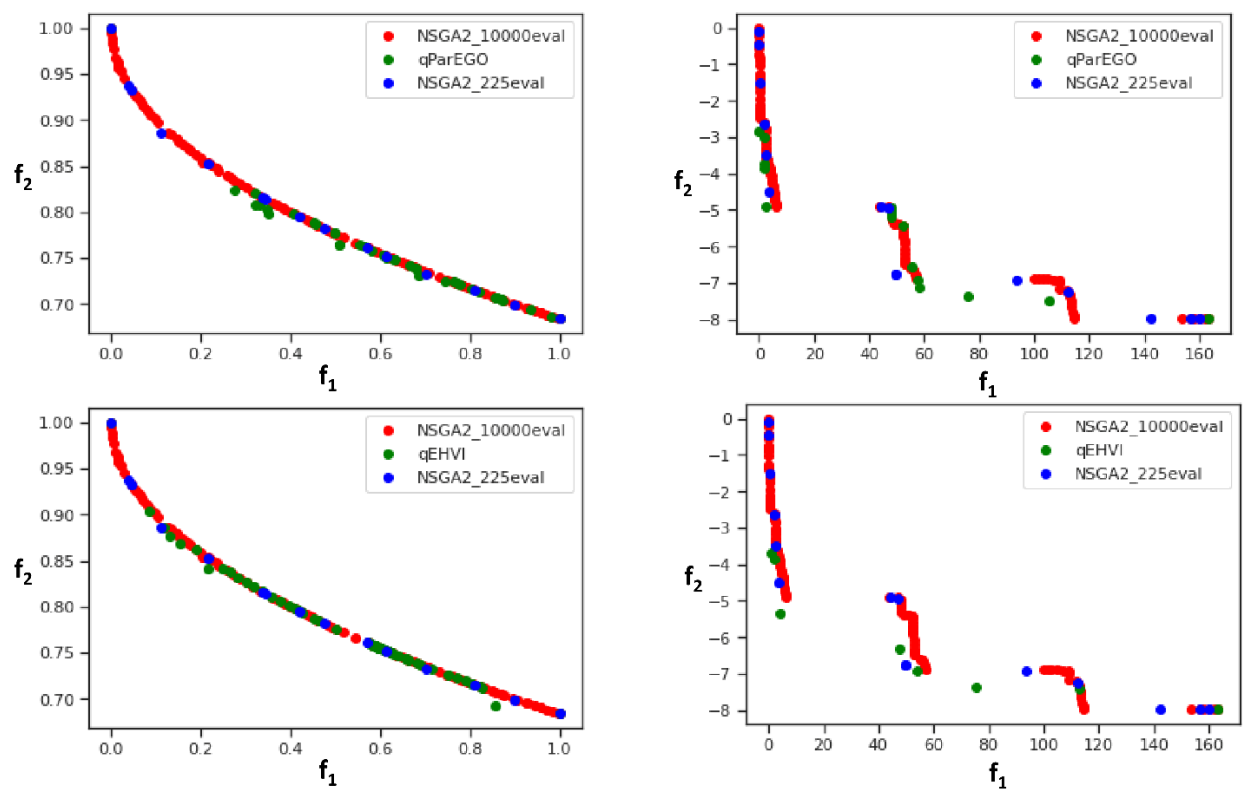

**Figure 7:** Comparison of Pareto-frontier between ground truth (using NSGA2), similar # of function evaluation in NSGA2 and MOBO for (a) ZDT1, acquisition function- qParEGO, (b) ZDT1, acquisition function- qEIHV, (c) 6HC-IAP, acquisition function- qParEGO, and (d) 6HC-IAP, acquisition function- qEIHV.

## 3. **MOBO of chemically-controlled ferroelectric**

With the general principles of MOBO discussed in Section 2 and interface-controlled ferroelectric model introduced in Section 1, here we present a MOBO optimization of the energy storage and losses in this system. For a given set of model parameters, we can generate the hysteresis loops for the field-dependent polarization response. The per se are too complex (high dimensionality) in the MOBO framework to define objective functions, and train the MOBO model. Thus, we build a decision tree to first extract different low dimensional target functions from different loop structures, given the numerical physics driven model parameters through an automated task.

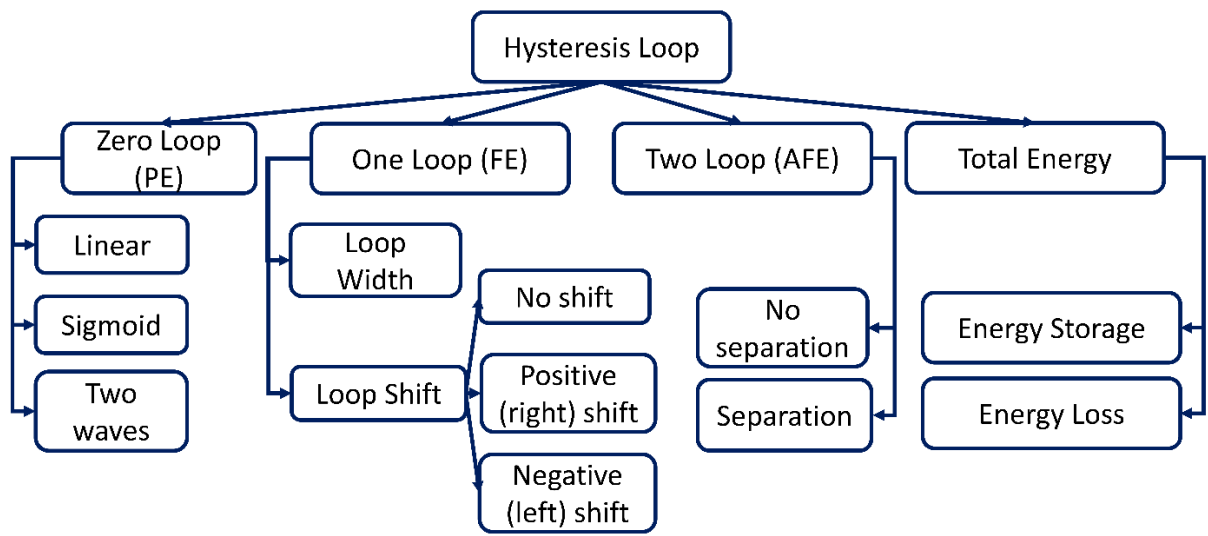

**Figure 8:** Physics driven decision tree from complex loop structures.

Figure 8 shows the full decision tree. For chosen parameters of a material (eg. 2-4-6 KLGD for bulk PZO), we can get different loop structures such as zero loop, one loop and two loops which defines different phases like paraelectric (PE), ferroelectric (FE) and anti-ferroelectric (AFE) respectively. For each of these phases, we have different classifications. For example, if the phase is PE, we then classify the hysteresis curve as linear, sigmoid and two waves. If the phase is AFE, we further classify as the two loops are attached or not, with quantifying the distance between separation between the loops. If the phase is FE, we have two different classifications. The first one is the measure of the shift of the loops, and in positive or negative direction. The second one is the measure of the width of the loop. Finally, as shown in fig. 8, irrespective of the loop structures, we calculate the energy storage and loss.

As the focus of this paper is the performance optimization for memory and energy storage applications, we selected the energy storage and loss as the objectives of the MOBO. It is to be noted that the stated automated task of developing a similar decision tree from the hysteresis loop can be applicable to any theoretical model or experimental settings of any materials and the user can choose any target functions (branch) in the decision tree as the objectives as appropriate for the optimization problem. Finally, integrating the decision tree in fig 8 with the general MOBO framework (Table 1), we develop a physics-driven MOBO architecture to optimize various target material properties jointly and thus attempt to rapidly explore multi-dimensional parameter space for initial material and device parameter selection. The full architecture is summarized in fig. 9. The MOBO framework is coded in Python BoTorch package. GP models are configured using GPyTorch package and to handle multiple independent GP models, we utilize multi-output GP function, ModelListGP. The full code is available in Colab notebook.

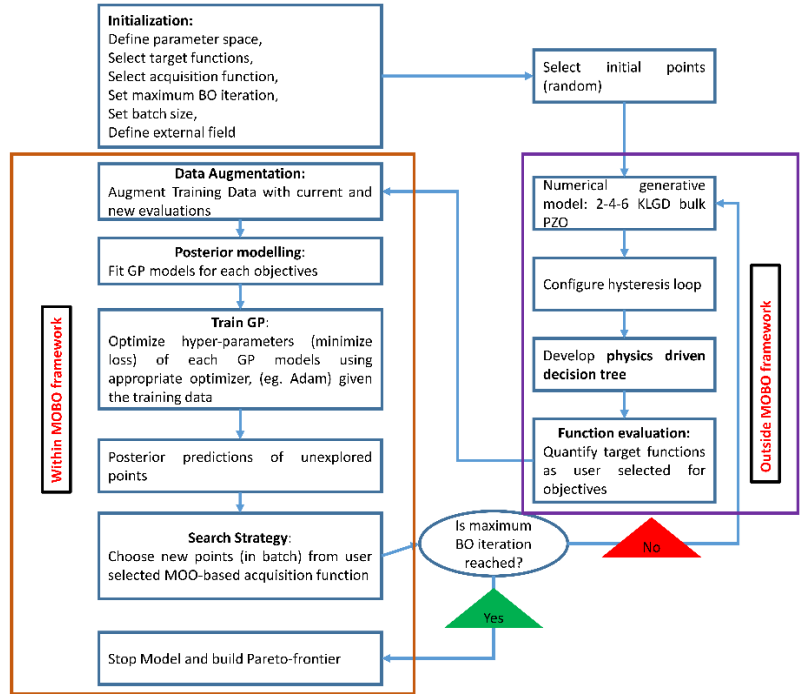

**Figure 9:** Full MOBO architecture: Application to physics driven numerical models to optimize target material properties.

Now, we illustrate the application the proposed physics-driven MOBO architecture in the 2-4-6 KLGD for bulk PZO which contains different phases such as PE, FE and AFE for chosen parameter space. Here, we aim to explore the balance between energy storage and loss, and build the Pareto front where we have various flavors of solutions – high storage-low loss, medium storage-medium loss, low storage-high loss. It is up to the decision maker to choose the appropriate pareto-optimal solution(s) which gives the desired balance in device parameter selection for given application. We started with 20 randomly selected initial samples and then stopped MOBO after 30 iterations, with 4 samples (batch_size =4) selected at each iteration (30x4=120 sampling). Figures 10-12 shows the Pareto-frontiers through rapid exploration-exploitation using MOBO with priori and posteriori acquisition functions over different parameter space such as such as Temperature, partial $O_2$ pressure, film thickness and surface ion energy.

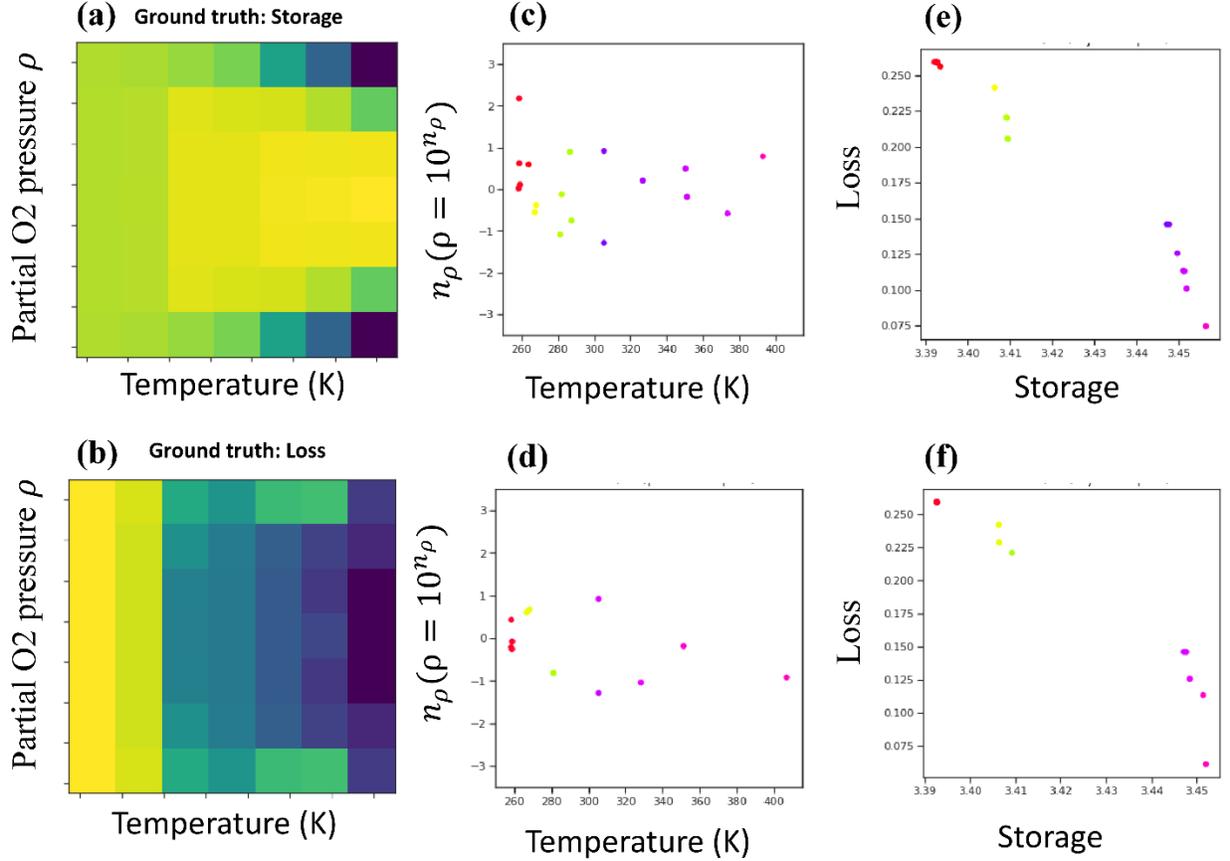

**Figure 10:** Parameter space $T = [250, 410]K, n_\rho = [-3, \ 3]$ where $\rho = 10^{n_\rho}, h = 5nm, \Delta_G = 0.2eV$. External field, $E = 3 * \sin(2\omega t)$. This parameter space falls into **AFE**. (a-b) Exhaustive low-sampling (7x7) grid exploration under the same parameter space, (a) Ground truth for function 1 (Storage), (b) Ground truth for function 2 (loss). (c-d) Pareto-optimals, explored with qParEGO and qEIHV acquisition functions respectively in the parameter space. (e-f) Pareto-frontiers drawn, after exploring with qParEGO and qEIHV acquisition functions respectively in the objective space. The colors map the projection of the pareto-solutions between the objective (e, f) and parameter space (c, d) respectively. The y-labels of the parameter space of figs a-d are same, for better visualization to compare with ground truth figures, we have drawn the pareto plots over the power of 10 of the values of partial $O_2$ pressure, $\rho$.

To check the performance of MOBO, we draw the ground truth of storage and loss (figs 10 a, b) using low sampling (7x7), expensive exhaustive (grid-based) exploration. Here, we configure the pareto-frontier within the AFE phase only. Interpreting the trade-off between storage and loss in the parameter and objective space (figs 10 c-f), we see that higher storage maps to higher temperature and lower order (in the absolute power of 10) partial $O_2$ pressure whereas lower storage maps generally to lower temperature and higher order partial $O_2$ pressure at the chosen value of film thickness and surface ion energy. The balanced trade-off between storage and loss populates the intermediate values in the parameter space. The ground truth shows the similar maps of high and low values of storage and loss in the parameter space. As we compare the pareto frontier and the ground truth in the parameter space (figs. 10 a-d), the pareto frontier actually connects the maximum value regions of both storage and loss, giving a trade-off between them. Also, we present the loop diagrams in appendix fig. 13, from where the ground truth of storage/loss is drawn following the decision tree. It is clearly seen that the loop area is decreasing as the temperature and partial $O_2$ pressure is higher and lower (in the absolute power of 10) respectively, thus maximizing the storage or minimizing the loss. This agrees to the results in fig. 10. Both the acquisition functions performed quite similar.

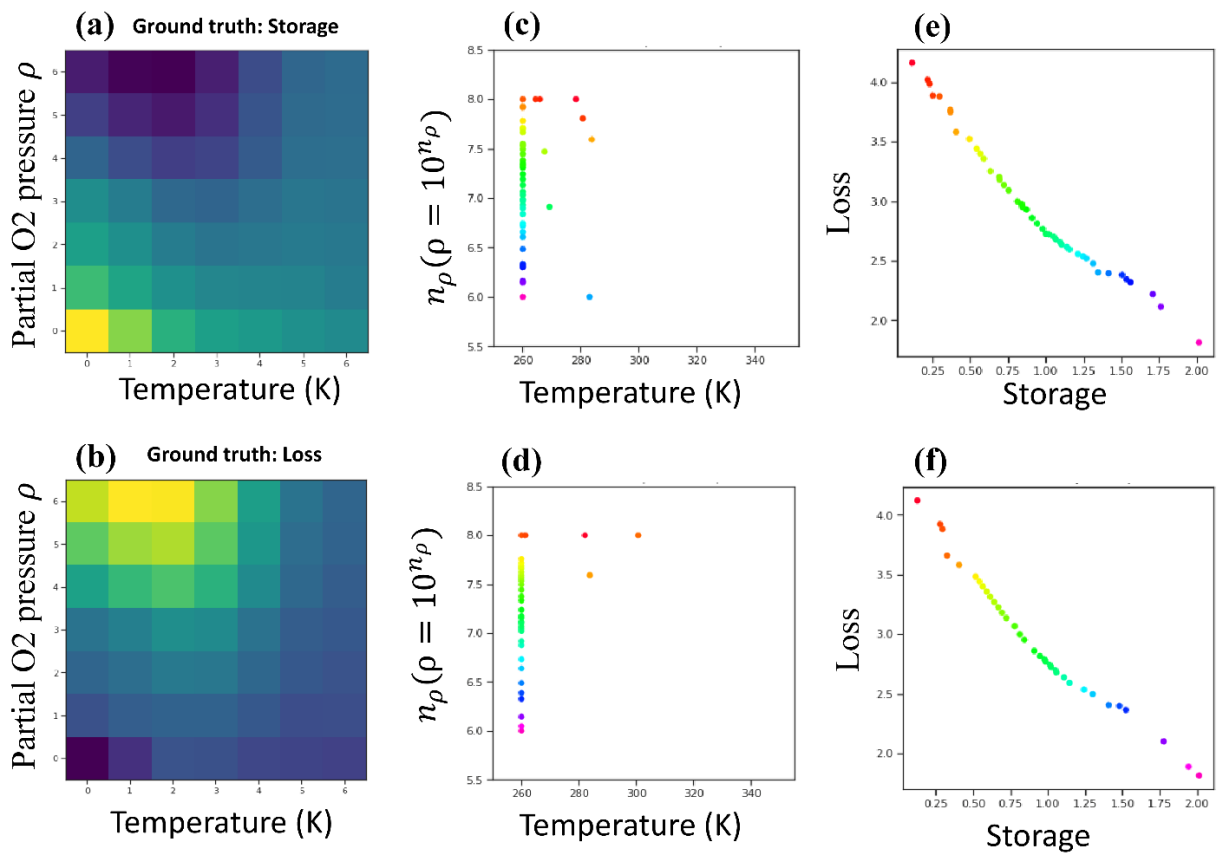

**Figure 11:** Parameter space $T = [260, 350]K$, $n_\rho = [6, 8]$ *where* $\rho = 10^{n_\rho}$, $h = 50nm$, $\Delta_G = 0.2eV$. External field, $E = 3 * \sin(2\omega t)$. This parameter space falls into **FE**. (a-b) Exhaustive low-sampling (7x7) grid exploration under the same parameter space, (a) Ground truth for function 1 (Storage), (b) Ground truth for function 2 (loss). (c-d) Pareto-optimals, explored with qParEGO and qEIHV acquisition functions respectively in the parameter space. (e-f) Pareto-frontiers drawn, after exploring with qParEGO and qEIHV acquisition functions respectively in the objective space. The colors map the projection of the pareto-solutions between the objective (e, f) and parameter space (c, d) respectively. The y-labels of the parameter space of figs a-d are same, for better visualization to compare with ground truth figures, we have drawn the pareto plots over the power of 10 of the values of partial $O_2$ pressure, $\rho$.

Also in fig 11, we first draw the ground truth of storage and loss (figs 11 a, b). Here, we configure the pareto-frontier within the FE phase only. Interpreting the trade-off between storage and loss in the parameter and objective space (figs 11 c-f), we see that higher loss maps to higher partial $O_2$ pressure whereas lower loss maps to lower partial $O_2$ pressure at the chosen value of film thickness and surface ion energy. The balanced trade-off between storage and loss populates the intermediate values of partial $O_2$ pressure. Though the dependence of temperature is much lesser than partial O₂ pressure in the effect of storage or loss in the FE phase at the given parameters, however, most of pareto solutions in the intermediate region are at low temperature. Validating the interpretation with the ground truth using low sampling (7x7), expensive exhaustive (grid-based) exploration, it shows the similar maps of high and low values of storage and loss in the parameter space. As we compare the pareto frontier and the ground truth in the parameter space (figs. 11 a-d), the pareto frontier actually connects the maximum value regions of both storage and loss, giving a trade-off between them. Here, also we see the dependence of partial O₂ pressure is much greater than temperature, which goes by the results from MOBO. From the loops diagrams in appendix fig. 14, it is clearly seen that the loop area is increasing as the partial O₂ pressure is higher, thus minimizing the storage or maximizing the loss. This agrees to the results in fig. 11. Both the acquisition functions performed quite similar.

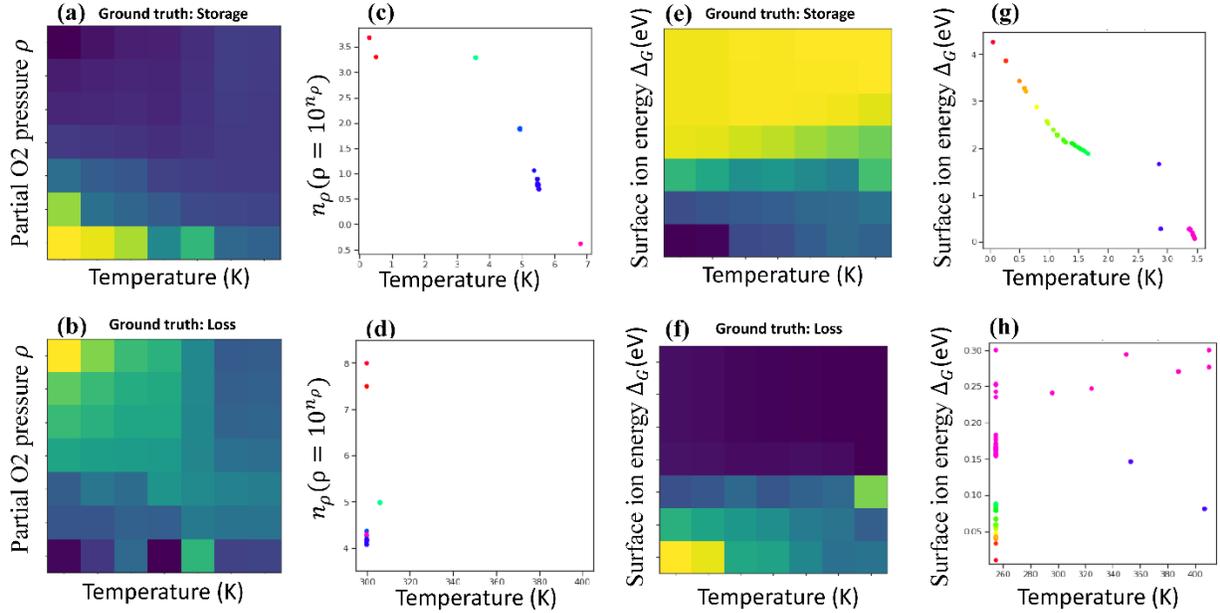

**Figure 12:** (a-d) Parameter space $T = [255, 410]K$, $\rho = [10^2, 10^8]$, $h = 50nm$, $\Delta_G = 0.2eV$. External field, $E = 3 * \sin(2\omega t)$. Exhaustive low-sampling (7x7) grid exploration under the same parameter space, (a) Ground truth for function 1 (Storage), (b) Ground truth for function 2 (loss). Same pareto-frontier, explored with qParEGO acquisition function in (c) objective space and in (d) parameter space. The y-labels of the parameter space of figs a-d are same, for better visualization to compare with ground truth figures, we have drawn the pareto plots over the power of 10 of the values of partial O₂ pressure, $\rho$. (e-h) Parameter space $T = [255, 410]K$, $\rho = 10^2$, $h = 5nm$, $\Delta_G = [0.01, 0.3]eV$. External field, $E = 3 * \sin(2\omega t)$. Exhaustive low-sampling (7x7) grid exploration under the same parameter space, (e) Ground truth for function 1 (Storage), (f) Ground truth for function 2 (loss). Same pareto-frontier, explored with qEIHV acquisition function in (g) objective space and in (h) parameter space. Both parameter spaces fall into both **AFE** and **FE**.

Both the parameter space in fig 12 configure the pareto-frontier within both AFE and FE phases. Focusing on the parameter space as in 12 a-d, we first draw the ground truth of storage and loss (figs 12 a, b). Interpreting the trade-off between storage and loss in the objective and parameter space (figs. 12 c, d), we see that higher loss maps to higher partial $O_2$ pressure and lower temperature whereas higher storage maps to lower partial $O_2$ pressure at the chosen value of film thickness and surface ion energy. Validating the interpretation with the ground truth using low sampling (7x7), expensive exhaustive (grid-based) exploration, it shows the similar maps of high and low values of storage and loss in the parameter space. As we compare the pareto frontier and the ground truth in the parameter space (12 a, b, d), the pareto frontier actually connects the maximum value regions of both storage and loss, giving a trade-off between them. From the loops diagrams in appendix fig. 15, it is clearly seen that the loop area (considering both single loop FE region and double loop AFE region) is increasing as the partial $O_2$ pressure is higher and temperature is mostly lower (as we move from AFE to FE, and deeper into FE domain) and is decreasing as the partial $O_2$ pressure is lower (as we move from FE to AFE, and deeper into AFE domain), thus minimizing (maximizing) the storage (loss) and maximizing (minimizing) the storage (loss) respectively. This agrees to the results in figs. 12 a-d. Interestingly, we see a sudden jump in storage and loss in the pareto after very closely space pareto solutions, which likely happened in this case, due to phase transition between FE and AFE (see appendix fig. 15).

Focusing on the parameter space (choosing temperature and surface ion energy) as in 12 e-h, we first draw the ground truth of storage and loss (figs 12 e, f). Interpreting the trade-off between storage and loss in the objective and parameter space (figs 12 g, h), we see that higher loss maps to lower surface ion energy and temperature whereas higher storage maps to higher surface ion energy at the chosen value of film thickness and partial $O_2$ pressure. Validating the interpretation with the ground truth using low sampling (7x7), expensive exhaustive (grid-based) exploration, we find similar maps of high and low values of storage and loss in the parameter space. As we compare the pareto frontier and the ground truth in the parameter space (12 e, f, h), the pareto frontier actually connects the maximum value regions of both storage and loss, giving a trade-off between them. From the loops diagrams in appendix fig. 16, it is clearly seen that the loop area (considering both single loop FE region and double loop AFE region) is increasing as the surface ion energy and temperature is lower (as we move from AFE to FE, and deeper into FE domain) and is decreasing as the surface ion energy is higher (as we move from FE to AFE, and deeper into AFE domain), thus minimizing (maximizing) the storage (loss) and maximizing (minimizing) the storage (loss) respectively. This agrees to the results in figs. 12 e-h. We see a similar jump of storage and loss, which again likely to be due to phase transition between FE and AFE. (see appendix fig. 16). In practical, we won't know this transition region between the phases and this type of sharp jumps can be problematic for the surrogate GP model in MOBO. Thus, to avoid this type of sudden jump at the unknown transition boundary, we can build pareto frontier (trade-off between the target functions) over any parameter space for a pre-selected phase (AFE or FE) using more complex architecture of constrained Multi-objective Bayesian optimization (MOBO), which will be addressed in future. More detailed figures for the analysis of 2-4-6 KLGD for bulk PZO, with the iteration (steps-wise) of MOBO is provided in Colab notebook.

## 4. Summary

To summarize, here we develop a workflow for the joint optimization of multiple materials functionalities from expensive computations using the multi objective Bayesian Optimization. This approach is applied for the discovery of optimal Pareto frontier between the energy storage and loss in interfacially-controlled ferroelectric materials but can be equally applied to other functionalities that can be derived from the P-E hysteresis loops.

In general, MOBO offers a powerful tool for the optimization of materials functionalities and can be further extended towards theory-experiment matching. As in practical problems, there is a general need to focus of different functionalities from expensive experiments, which is more complex in the field of optimization. Here, the proposed MOBO architecture offers a great choice where it first replaces with cheap surrogate theoretical modeling with physics driven prior knowledge and then optimizes any chosen multiple functionalities jointly in attempt to rapidly discover the optimal systems with desired balance between the functionalities. Going forward, this approach can help in guiding the experiments by exploring towards the region of interest in high-dimensional multi-functionality space, and thereby reducing time and effort.

**Acknowledgements:**

This work was supported (A.B.) by the US Department of Energy, Office of Science, Office of Basic Energy Sciences, as part of the Energy Frontier Research Centers program: CSSAS–The Center for the Science of Synthesis Across Scales–under Award Number DE-SC0019288, located at University of Washington, by (S.V.K.) the center for 3D Ferroelectric Microelectronics (3DFeM), an Energy Frontier Research Center funded by the U.S. Department of Energy (DOE), Office of Science, Basic Energy Sciences under Award Number DE-SC0021118 and performed at Oak Ridge National Laboratory's Center for Nanophase Materials Sciences (CNMS), a U.S. Department of Energy, Office of Science User Facility. A.N.M work is supported by the National Research Foundation of Ukraine (Grant application Φ81/41481).

**Author Declarations:**

- **Conflict of interest:**
    - The authors have no conflicts to disclose.

**Data Availability:**

The general findings in this manuscript are summarized in the notebooks at https://github.com/arpanbiswas52/MOBO_AFI_Supplements.

# References

(1)     Tagantsev, A.; Cross, L. E.; Fousek, J. *Domains in Ferroic Crystals and Thin Films*; Springer: New York, 2010.

(2)     Auciello, O.; Scott, J. F.; Ramesh, R. The Physics of Ferroelectric Memories. *Physics Today* **1998**, *51* (7), 22–27. https://doi.org/10.1063/1.882324.

(3)     Scott, J. F.; Araujo, C. A. P. de. Ferroelectric Memories. *Science* **1989**, *246* (4936), 1400–1405. https://doi.org/10.1126/science.246.4936.1400.

(4)     Tsymbal, E. Y.; Kohlstedt, H. Tunneling Across a Ferroelectric. *Science* **2006**, *313* (5784), 181–183. https://doi.org/10.1126/science.1126230.

(5)     Gruverman, A.; Wu, D.; Lu, H.; Wang, Y.; Jang, H. W.; Folkman, C. M.; Zhuravlev, M. Ye.; Felker, D.; Rzchowski, M.; Eom, C.-B.; Tsymbal, E. Y. Tunneling Electroresistance Effect in Ferroelectric Tunnel Junctions at the Nanoscale. *Nano Lett.* **2009**, *9* (10), 3539–3543. https://doi.org/10.1021/nl901754t.

(6)     Maksymovych, P.; Jesse, S.; Yu, P.; Ramesh, R.; Baddorf, A. P.; Kalinin, S. V. Polarization Control of Electron Tunneling into Ferroelectric Surfaces. *Science* **2009**, *324* (5933), 1421–1425. https://doi.org/10.1126/science.1171200.

(7)     Miller, S. L.; McWhorter, P. J. Physics of the Ferroelectric Nonvolatile Memory Field Effect Transistor. *Journal of Applied Physics* **1992**, *72* (12), 5999–6010. https://doi.org/10.1063/1.351910.

(8)     Mathews, S.; Ramesh, R.; Venkatesan, T.; Benedetto, J. Ferroelectric Field Effect Transistor Based on Epitaxial Perovskite Heterostructures. *Science* **1997**, *276* (5310), 238–240. https://doi.org/10.1126/science.276.5310.238.

(9)     Garcia, V.; Bibes, M.; Bocher, L.; Valencia, S.; Kronast, F.; Crassous, A.; Moya, X.; Enouz-Vedrenne, S.; Gloter, A.; Imhoff, D.; Deranlot, C.; Mathur, N. D.; Fusil, S.; Bouzehouane, K.; Barthélémy, A. Ferroelectric Control of Spin Polarization. *Science* **2010**, *327* (5969), 1106–1110. https://doi.org/10.1126/science.1184028.

(10)    Catalan, G.; Scott, J. F. Physics and Applications of Bismuth Ferrite. *Advanced Materials* **2009**, *21* (24), 2463–2485. https://doi.org/10.1002/adma.200802849.

(11)    Balashova, E. V.; Tagantsev, A. K. Polarization Response of Crystals with Structural and Ferroelectric Instabilities. *Phys. Rev. B* **1993**, *48* (14), 9979–9986. https://doi.org/10.1103/PhysRevB.48.9979.

(12)    Balashova, E. V.; Lemanov, V. V.; Tagantsev, A. K.; Sherman, A. B.; Shomuradov, Sh. H. Betaine Arsenate as a System with Two Instabilities. *Phys. Rev. B* **1995**, *51* (14), 8747–8752. https://doi.org/10.1103/PhysRevB.51.8747.

(13)    Chensky, E. V.; Tarasenko, V. V. Theory of Phase-Transitions to Inhomogeneous States in Finite Ferroelectrics in an External Electric-Field. *Sov. Phys. JETP., 56: 618 [Zh. Eksp. Teor. Fiz. 83, 1089 (1982)]*.

(14)    Bratkovsky, A. M.; Levanyuk, A. P. Continuous Theory of Ferroelectric States in Ultrathin Films with Real Electrodes. *Journal of Computational and Theoretical Nanoscience* **2009**, *6* (3), 465–489. https://doi.org/10.1166/jctn.2009.1058.

(15)    Bratkovsky, A. M.; Levanyuk, A. P. Effects of Anisotropic Elasticity in the Problem of Domain Formation and Stability of Monodomain State in Ferroelectric Films. *Phys. Rev. B* **2011**, *84* (4), 045401. https://doi.org/10.1103/PhysRevB.84.045401.

(16)    Chen, L. Q.; Khachaturyan, A. G. Dynamics of Simultaneous Ordering and Phase Separation and Effect of Long-Range Coulomb Interactions. *Phys. Rev. Lett.* **1993**, *70* (10), 1477–1480. https://doi.org/10.1103/PhysRevLett.70.1477.

(17)  Fan, D.; Chen, L.-Q. Possibility of Spinodal Decomposition in ZrO2-Y2O3 Alloys: A Theoretical Investigation. *Journal of the American Ceramic Society* **1995**, *78* (6), 1680–1686. https://doi.org/10.1111/j.1151-2916.1995.tb08870.x.

(18)  Balke, N.; Winchester, B.; Ren, W.; Chu, Y. H.; Morozovska, A. N.; Eliseev, E. A.; Huijben, M.; Vasudevan, R. K.; Maksymovych, P.; Britson, J.; Jesse, S.; Kornev, I.; Ramesh, R.; Bellaiche, L.; Chen, L. Q.; Kalinin, S. V. Enhanced Electric Conductivity at Ferroelectric Vortex Cores in BiFeO3. *Nature Phys* **2012**, *8* (1), 81–88. https://doi.org/10.1038/nphys2132.

(19)  Böscke, T. S.; Teichert, St.; Bräuhaus, D.; Müller, J.; Schröder, U.; Böttger, U.; Mikolajick, T. Phase Transitions in Ferroelectric Silicon Doped Hafnium Oxide. *Appl. Phys. Lett.* **2011**, *99* (11), 112904. https://doi.org/10.1063/1.3636434.

(20)  Müller, J.; Böscke, T. S.; Bräuhaus, D.; Schröder, U.; Böttger, U.; Sundqvist, J.; Kücher, P.; Mikolajick, T.; Frey, L. Ferroelectric Zr0.5Hf0.5O2 Thin Films for Nonvolatile Memory Applications. *Appl. Phys. Lett.* **2011**, *99* (11), 112901. https://doi.org/10.1063/1.3636417.

(21)  Park, M. H.; Chung, C.-C.; Schenk, T.; Richter, C.; Hoffmann, M.; Wirth, S.; Jones, J. L.; Mikolajick, T.; Schroeder, U. Origin of Temperature-Dependent Ferroelectricity in Si-Doped HfO2. *Advanced Electronic Materials* **2018**, *4* (4), 1700489. https://doi.org/10.1002/aelm.201700489.

(22)  Cao, J.; Shi, S.; Zhu, Y.; Chen, J. An Overview of Ferroelectric Hafnia and Epitaxial Growth. *physica status solidi (RRL) – Rapid Research Letters* **2021**, *15* (5), 2100025. https://doi.org/10.1002/pssr.202100025.

(23)  Lomenzo, P. D.; Schroeder, U.; Mikolajick, T. Electronic Contributions to Ferroelectricity and Field-Induced Phase Transitions in Doped-HfO2. In *2021 IEEE International Symposium on Applications of Ferroelectrics (ISAF)*; 2021; pp 1–4. https://doi.org/10.1109/ISAF51943.2021.9477323.

(24)  Zhu, W.; Hayden, J.; He, F.; Yang, J.-I.; Tipsawat, P.; Hossain, M. D.; Maria, J.-P.; Trolier-McKinstry, S. Strongly Temperature Dependent Ferroelectric Switching in AlN, Al1-XScxN, and Al1-XBxN Thin Films. *Appl. Phys. Lett.* **2021**, *119* (6), 062901. https://doi.org/10.1063/5.0057869.

(25)  Song, Y.; Perez, C.; Esteves, G.; Lundh, J. S.; Saltonstall, C. B.; Beechem, T. E.; Yang, J. I.; Ferri, K.; Brown, J. E.; Tang, Z.; Maria, J.-P.; Snyder, D. W.; Olsson, R. H.; Griffin, B. A.; Trolier-McKinstry, S. E.; Foley, B. M.; Choi, S. Thermal Conductivity of Aluminum Scandium Nitride for 5G Mobile Applications and Beyond. *ACS Appl. Mater. Interfaces* **2021**, *13* (16), 19031–19041. https://doi.org/10.1021/acsami.1c02912.

(26)  Ferri, K.; Bachu, S.; Zhu, W.; Imperatore, M.; Hayden, J.; Alem, N.; Giebink, N.; Trolier-McKinstry, S.; Maria, J.-P. Ferroelectrics Everywhere: Ferroelectricity in Magnesium Substituted Zinc Oxide Thin Films. *Journal of Applied Physics* **2021**, *130* (4), 044101. https://doi.org/10.1063/5.0053755.

(27)  Sang, X.; Grimley, E. D.; Schenk, T.; Schroeder, U.; LeBeau, J. M. On the Structural Origins of Ferroelectricity in HfO2 Thin Films. *Appl. Phys. Lett.* **2015**, *106* (16), 162905. https://doi.org/10.1063/1.4919135.

(28)  Delodovici, F.; Barone, P.; Picozzi, S. Phys. Rev. Mater. **2021**, *5* (6).

(29)  Nukala, P.; Ahmadi, M.; Wei, Y.; Graaf, S. de; Stylianidis, E.; Chakrabortty, T.; Matzen, S.; Zandbergen, H. W.; Björling, A.; Mannix, D.; Carbone, D.; Kooi, B.; Noheda, B. Reversible Oxygen Migration and Phase Transitions in Hafnia-Based Ferroelectric Devices. *Science* **2021**, *372* (6542), 630–635. https://doi.org/10.1126/science.abf3789.

(30)  Dogan, M.; Gong, N.; Ma, T.-P.; Ismail-Beigi, S. Causes of Ferroelectricity in HfO2-Based Thin Films: An Ab Initio Perspective. *Phys. Chem. Chem. Phys.* **2019**, *21* (23), 12150–12162. https://doi.org/10.1039/C9CP01880H.

(31)  Huan, T. D.; Sharma, V.; Rossetti, G. A.; Ramprasad, R. Pathways towards Ferroelectricity in Hafnia. *Phys. Rev. B* **2014**, *90* (6), 064111. https://doi.org/10.1103/PhysRevB.90.064111.

(32)     Wu, P.; Ma, X.; Li, Y.; Gopalan, V.; Chen, L.-Q. Dipole Spring Ferroelectrics in Superlattice SrTiO3/BaTiO3 Thin Films Exhibiting Constricted Hysteresis Loops. *Appl. Phys. Lett.* **2012**, *100* (9), 092905. https://doi.org/10.1063/1.3691172.

(33)     Scott, J. F.; Salje, E. K. H.; Carpenter, M. A. Domain Wall Damping and Elastic Softening in ${\mathrm{SrTiO}}_{3}$: Evidence for Polar Twin Walls. *Phys. Rev. Lett.* **2012**, *109* (18), 187601. https://doi.org/10.1103/PhysRevLett.109.187601.

(34)     Lee, Y.; Goh, Y.; Hwang, J.; Das, D.; Jeon, S. The Influence of Top and Bottom Metal Electrodes on Ferroelectricity of Hafnia. *IEEE Transactions on Electron Devices* **2021**, *68* (2), 523–528. https://doi.org/10.1109/TED.2020.3046173.

(35)     Morozovska, A. N.; Eliseev, E. A.; Svechnikov, S. V.; Krutov, A. D.; Shur, V. Y.; Borisevich, A. Y.; Maksymovych, P.; Kalinin, S. V. Finite Size and Intrinsic Field Effect on the Polar-Active Properties of Ferroelectric-Semiconductor Heterostructures. *Phys. Rev. B* **2010**, *81* (20), 205308. https://doi.org/10.1103/PhysRevB.81.205308.

(36)     Spanier, J. E.; Kolpak, A. M.; Urban, J. J.; Grinberg, I.; Ouyang, L.; Yun, W. S.; Rappe, A. M.; Park, H. Ferroelectric Phase Transition in Individual Single-Crystalline BaTiO3 Nanowires. *Nano Lett.* **2006**, *6* (4), 735–739. https://doi.org/10.1021/nl052538e.

(37)     Wang, R. V.; Fong, D. D.; Jiang, F.; Highland, M. J.; Fuoss, P. H.; Thompson, C.; Kolpak, A. M.; Eastman, J. A.; Streiffer, S. K.; Rappe, A. M.; Stephenson, G. B. Reversible Chemical Switching of a Ferroelectric Film. *Phys. Rev. Lett.* **2009**, *102* (4), 047601. https://doi.org/10.1103/PhysRevLett.102.047601.

(38)     Fong, D. D.; Kolpak, A. M.; Eastman, J. A.; Streiffer, S. K.; Fuoss, P. H.; Stephenson, G. B.; Thompson, C.; Kim, D. M.; Choi, K. J.; Eom, C. B.; Grinberg, I.; Rappe, A. M. Stabilization of Monodomain Polarization in Ultrathin ${\mathrm{PbTiO}}_{3}$ Films. *Phys. Rev. Lett.* **2006**, *96* (12), 127601. https://doi.org/10.1103/PhysRevLett.96.127601.

(39)     Kalinin, S. V.; Kim, Y.; Fong, D. D.; Morozovska, A. N. Surface-Screening Mechanisms in Ferroelectric Thin Films and Their Effect on Polarization Dynamics and Domain Structures. *Rep. Prog. Phys.* **2018**, *81* (3), 036502. https://doi.org/10.1088/1361-6633/aa915a.

(40)     Domingo, N.; Gaponenko, I.; Cordero-Edwards, K.; Stucki, N.; Pérez-Dieste, V.; Escudero, C.; Pach, E.; Verdaguer, A.; Paruch, P. Surface Charged Species and Electrochemistry of Ferroelectric Thin Films. *Nanoscale* **2019**, *11* (38), 17920–17930. https://doi.org/10.1039/C9NR05526F.

(41)     Glinchuk, M. D.; Morozovska, A. N.; Lukowiak, A.; Stręk, W.; Silibin, M. V.; Karpinsky, D. V.; Kim, Y.; Kalinin, S. V. Possible Electrochemical Origin of Ferroelectricity in HfO2 Thin Films. *Journal of Alloys and Compounds* **2020**, *830*, 153628. https://doi.org/10.1016/j.jallcom.2019.153628.

(42)     Zhou, C.; Newns, D. M. Intrinsic Dead Layer Effect and the Performance of Ferroelectric Thin Film Capacitors. *Journal of Applied Physics* **1997**, *82* (6), 3081–3088. https://doi.org/10.1063/1.366147.

(43)     Gerra, G.; Tagantsev, A. K.; Setter, N. Ferroelectricity in Asymmetric Metal-Ferroelectric-Metal Heterostructures: A Combined First-Principles--Phenomenological Approach. *Phys. Rev. Lett.* **2007**, *98* (20), 207601. https://doi.org/10.1103/PhysRevLett.98.207601.

(44)     Zubko, P.; Gariglio, S.; Gabay, M.; Ghosez, P.; Triscone, J.-M. Interface Physics in Complex Oxide Heterostructures. *Annu. Rev. Condens. Matter Phys.* **2011**, *2* (1), 141–165. https://doi.org/10.1146/annurev-conmatphys-062910-140445.

(45)     Bardeen, J. Surface States and Rectification at a Metal Semi-Conductor Contact. *Phys. Rev.* **1947**, *71* (10), 717–727. https://doi.org/10.1103/PhysRev.71.717.

(46)     Vorotiahin, I. S.; Eliseev, E. A.; Li, Q.; Kalinin, S. V.; Genenko, Y. A.; Morozovska, A. N. Tuning the Polar States of Ferroelectric Films via Surface Charges and Flexoelectricity. *Acta Materialia* **2017**, *137*, 85–92. https://doi.org/10.1016/j.actamat.2017.07.033.

(47)    Morozovska, A. N.; Eliseev, E. A.; Vorotiahin, I. S.; Silibin, M. V.; Kalinin, S. V.; Morozovsky, N. V. Control of Polarization Reversal Temperature Behavior by Surface Screening in Thin Ferroelectric Films. *Acta Materialia* **2018**, *160*, 57–71. https://doi.org/10.1016/j.actamat.2018.08.041.

(48)    Fridkin, V. M. *Ferroelectric Semiconductors.*; Springer, New York, 1980.

(49)    Stephenson, G. B.; Highland, M. J. Equilibrium and Stability of Polarization in Ultrathin Ferroelectric Films with Ionic Surface Compensation. *Phys. Rev. B* **2011**, *84* (6), 064107. https://doi.org/10.1103/PhysRevB.84.064107.

(50)    Highland, M. J.; Fister, T. T.; Fong, D. D.; Fuoss, P. H.; Thompson, C.; Eastman, J. A.; Streiffer, S. K.; Stephenson, G. B. Equilibrium Polarization of Ultrathin ${\mathrm{PbTiO}}_{3}$ with Surface Compensation Controlled by Oxygen Partial Pressure. *Phys. Rev. Lett.* **2011**, *107* (18), 187602. https://doi.org/10.1103/PhysRevLett.107.187602.

(51)    Yang, S. M.; Morozovska, A. N.; Kumar, R.; Eliseev, E. A.; Cao, Y.; Mazet, L.; Balke, N.; Jesse, S.; Vasudevan, R. K.; Dubourdieu, C.; Kalinin, S. V. Mixed Electrochemical–Ferroelectric States in Nanoscale Ferroelectrics. *Nature Phys* **2017**, *13* (8), 812–818. https://doi.org/10.1038/nphys4103.

(52)    Morozovska, A. N.; Eliseev, E. A.; Morozovsky, N. V.; Kalinin, S. V. Ferroionic States in Ferroelectric Thin Films. *Phys. Rev. B* **2017**, *95* (19), 195413. https://doi.org/10.1103/PhysRevB.95.195413.

(53)    Morozovska, A. N.; Eliseev, E. A.; Morozovsky, N. V.; Kalinin, S. V. Piezoresponse of Ferroelectric Films in Ferroionic States: Time and Voltage Dynamics. *Appl. Phys. Lett.* **2017**, *110* (18), 182907. https://doi.org/10.1063/1.4979824.

(54)    Morozovska, A. N.; Eliseev, E. A.; Kurchak, A. I.; Morozovsky, N. V.; Vasudevan, R. K.; Strikha, M. V.; Kalinin, S. V. Effect of Surface Ionic Screening on the Polarization Reversal Scenario in Ferroelectric Thin Films: Crossover from Ferroionic to Antiferroionic States. *Phys. Rev. B* **2017**, *96* (24), 245405. https://doi.org/10.1103/PhysRevB.96.245405.

(55)    Morozovska, A. N.; Eliseev, E. A.; Biswas, A.; Morozovsky, N. V.; Kalinin, S. V. Effect of Surface Ionic Screening on Polarization Reversal and Phase Diagrams in Thin Antiferroelectric Films for Information and Energy Storage. *arXiv:2106.13096 [cond-mat]* **2021**.

(56)    Tagantsev, A. K.; Gerra, G. Interface-Induced Phenomena in Polarization Response of Ferroelectric Thin Films. *Journal of Applied Physics* **2006**, *100* (5), 051607. https://doi.org/10.1063/1.2337009.

(57)    Blinc, R.; Zeks, B. *Soft Mode in Ferroelectrics and Antiferroelectrics*; North-Holland Publishing Company, Amsterdam, Oxford, 1974.

(58)    Tagantsev, A. K.; Vaideeswaran, K.; Vakhrushev, S. B.; Filimonov, A. V.; Burkovsky, R. G.; Shaganov, A.; Andronikova, D.; Rudskoy, A. I.; Baron, A. Q. R.; Uchiyama, H.; Chernyshov, D.; Bosak, A.; Ujma, Z.; Roleder, K.; Majchrowski, A.; Ko, J.-H.; Setter, N. The Origin of Antiferroelectricity in PbZrO3. *Nat Commun* **2013**, *4* (1), 2229. https://doi.org/10.1038/ncomms3229.

(59)    Rabe, K. M. Antiferroelectricity in Oxides: A Reexamination. In *Functional Metal Oxides*; John Wiley & Sons, Ltd; pp 221–244. https://doi.org/10.1002/9783527654864.ch7.

(60)    Hlinka, J.; Ostapchuk, T.; Buixaderas, E.; Kadlec, C.; Kuzel, P.; Gregora, I.; Kroupa, J.; Savinov, M.; Klic, A.; Drahokoupil, J.; Etxebarria, I.; Dec, J. Multiple Soft-Mode Vibrations of Lead Zirconate. *Phys. Rev. Lett.* **2014**, *112* (19), 197601. https://doi.org/10.1103/PhysRevLett.112.197601.

(61)    Foo, K. Y.; Hameed, B. H. Insights into the Modeling of Adsorption Isotherm Systems. *Chemical Engineering Journal* **2010**, *156* (1), 2–10. https://doi.org/10.1016/j.cej.2009.09.013.

(62)    Tian, Y.; Wei, L.; Zhang, Q.; Huang, H.; Zhang, Y.; Zhou, H.; Ma, F.; Gu, L.; Meng, S.; Chen, L.-Q.; Nan, C.-W.; Zhang, J. Water Printing of Ferroelectric Polarization. *Nat Commun* **2018**, *9* (1), 3809. https://doi.org/10.1038/s41467-018-06369-w.

(63)  Cao, Y.; Kalinin, S. V. Phase-Field Modeling of Chemical Control of Polarization Stability and Switching Dynamics in Ferroelectric Thin Films. *Phys. Rev. B* **2016**, *94* (23), 235444. https://doi.org/10.1103/PhysRevB.94.235444.

(64)  Sze, S. M. *Physics of Semiconductor Devices: 2nd Ed*, 2nd edition.; John Wiley and Sons Ltd: New York, N.Y., 1981.

(65)  Brochu, E.; Cora, V. M.; de Freitas, N. A Tutorial on Bayesian Optimization of Expensive Cost Functions, with Application to Active User Modeling and Hierarchical Reinforcement Learning. *arXiv:1012.2599 [cs]* **2010**.

(66)  Lizotte, D.; Wang, T.; Bowling, M.; Schuurmans, D. Automatic Gait Optimization with Gaussian Process Regression.; 2007; pp 944–949.

(67)  Lizotte, D. Practical Bayesian Optimization. **2008**.

(68)  Cora, V. M. Model-Based Active Learning in Hierarchical Policies, University of British Columbia, 2008. https://doi.org/10.14288/1.0051276.

(69)  Frean, M.; Boyle, P. Using Gaussian Processes to Optimize Expensive Functions. In *AI 2008: Advances in Artificial Intelligence*; Wobcke, W., Zhang, M., Eds.; Lecture Notes in Computer Science; Springer: Berlin, Heidelberg, 2008; pp 258–267. https://doi.org/10.1007/978-3-540-89378-3_25.

(70)  Martinez-Cantin, R.; de Freitas, N.; Brochu, E.; Castellanos, J.; Doucet, A. A Bayesian Exploration-Exploitation Approach for Optimal Online Sensing and Planning with a Visually Guided Mobile Robot. *Auton Robot* **2009**, *27* (2), 93–103. https://doi.org/10.1007/s10514-009-9130-2.

(71)  Chu, W.; Ghahramani, Z. Extensions of Gaussian Processes for Ranking: Semisupervised and Active Learning. *Learning to Rank* **2005**, 29.

(72)  Thurstone, L. L. A Law of Comparative Judgment. *Psychological Review* **1927**, *34* (4), 273–286. https://doi.org/10.1037/h0070288.

(73)  Mosteller, F. Remarks on the Method of Paired Comparisons: I. The Least Squares Solution Assuming Equal Standard Deviations and Equal Correlations. In *Selected Papers of Frederick Mosteller*; Fienberg, S. E., Hoaglin, D. C., Eds.; Springer Series in Statistics; Springer: New York, NY, 2006; pp 157–162. https://doi.org/10.1007/978-0-387-44956-2_8.

(74)  Holmes, C. C.; Held, L. Bayesian Auxiliary Variable Models for Binary and Multinomial Regression. *Bayesian Anal.* **2006**, *1* (1), 145–168. https://doi.org/10.1214/06-BA105.

(75)  Kotthoff, L.; Wahab, H.; Johnson, P. Bayesian Optimization in Materials Science: A Survey. *arXiv:2108.00002 [cond-mat, physics:physics]* **2021**.

(76)  Kiyohara, S.; Oda, H.; Tsuda, K.; Mizoguchi, T. Acceleration of Stable Interface Structure Searching Using a Kriging Approach. *Jpn. J. Appl. Phys.* **2016**, *55* (4), 045502. https://doi.org/10.7567/JJAP.55.045502.

(77)  Kikuchi, S.; Oda, H.; Kiyohara, S.; Mizoguchi, T. Bayesian Optimization for Efficient Determination of Metal Oxide Grain Boundary Structures. *Physica B: Condensed Matter* **2018**, *532*, 24–28. https://doi.org/10.1016/j.physb.2017.03.006.

(78)  Ueno, T.; Rhone, T. D.; Hou, Z.; Mizoguchi, T.; Tsuda, K. COMBO: An Efficient Bayesian Optimization Library for Materials Science. *Materials Discovery* **2016**, *4*, 18–21. https://doi.org/10.1016/j.md.2016.04.001.

(79)  Talapatra, A.; Boluki, S.; Duong, T.; Qian, X.; Dougherty, E.; Arróyave, R. Autonomous Efficient Experiment Design for Materials Discovery with Bayesian Model Averaging. *Phys. Rev. Materials* **2018**, *2* (11), 113803. https://doi.org/10.1103/PhysRevMaterials.2.113803.

(80)  Hankins, S.; Kotthoff, L.; Ray S. Fertig, I. Bio-like Composite Microstructure Designs for Enhanced Damage Tolerance via Machine Learning. *Proceedings of the American Society for Composites — Thirty-fourth Technical Conference* **2019**, *0* (0). https://doi.org/10.12783/asc34/31323.

(81)  Kalinin, S. V.; Ziatdinov, M.; Vasudevan, R. K. Guided Search for Desired Functional Responses via Bayesian Optimization of Generative Model: Hysteresis Loop Shape Engineering in Ferroelectrics. *Journal of Applied Physics* **2020**, *128* (2), 024102. https://doi.org/10.1063/5.0011917.

(82)  Ren, F.; Ward, L.; Williams, T.; Laws, K. J.; Wolverton, C.; Hattrick-Simpers, J.; Mehta, A. Accelerated Discovery of Metallic Glasses through Iteration of Machine Learning and High-Throughput Experiments. *Science Advances* **2018**, *4* (4), eaaq1566. https://doi.org/10.1126/sciadv.aaq1566.

(83)  Häse, F.; Roch, L. M.; Kreisbeck, C.; Aspuru-Guzik, A. Phoenics: A Bayesian Optimizer for Chemistry. *ACS Cent. Sci.* **2018**, *4* (9), 1134–1145. https://doi.org/10.1021/acscentsci.8b00307.

(84)  Kotthoff, L.; Jain, V.; Tyrrell, A.; Wahab, H.; Johnson, P. AI for Materials Science: Tuning Laser-Induced Graphene Production. In *In Data Science Meets Optimisation Workshop*; 2019.

(85)  Gopakumar, A. M.; Balachandran, P. V.; Xue, D.; Gubernatis, J. E.; Lookman, T. Multi-Objective Optimization for Materials Discovery via Adaptive Design. *Sci Rep* **2018**, *8* (1), 3738. https://doi.org/10.1038/s41598-018-21936-3.

(86)  Solomou, A.; Zhao, G.; Boluki, S.; Joy, J. K.; Qian, X.; Karaman, I.; Arróyave, R.; Lagoudas, D. C. Multi-Objective Bayesian Materials Discovery: Application on the Discovery of Precipitation Strengthened NiTi Shape Memory Alloys through Micromechanical Modeling. *Materials & Design* **2018**, *160*, 810–827. https://doi.org/10.1016/j.matdes.2018.10.014.

(87)  Hutter, F.; Hoos, H. H.; Leyton-Brown, K. Sequential Model-Based Optimization for General Algorithm Configuration. In *Learning and Intelligent Optimization*; Coello, C. A. C., Ed.; Lecture Notes in Computer Science; Springer: Berlin, Heidelberg, 2011; pp 507–523. https://doi.org/10.1007/978-3-642-25566-3_40.

(88)  Shahriari, B.; Swersky, K.; Wang, Z.; Adams, R. P.; de Freitas, N. Taking the Human Out of the Loop: A Review of Bayesian Optimization. *Proceedings of the IEEE* **2016**, *104* (1), 148–175. https://doi.org/10.1109/JPROC.2015.2494218.

(89)  Andrianakis, I.; Challenor, P. The Effect of the Nugget on Gaussian Process Emulators of Computer Models. *Computational Statistics & Data Analysis* **2012**, *56*, 4215–4228. https://doi.org/10.1016/j.csda.2012.04.020.

(90)  Pepelyshev, A. The Role of the Nugget Term in the Gaussian Process Method. In *mODa 9 − Advances in Model-Oriented Design and Analysis*; Giovagnoli, A., Atkinson, A. C., Torsney, B., May, C., Eds.; Contributions to Statistics; Physica-Verlag HD: Heidelberg, 2010; pp 149–156. https://doi.org/10.1007/978-3-7908-2410-0_20.

(91)  Xing, W.; Elhabian, S. Y.; Keshavarzzadeh, V.; Kirby, R. M. Shared-Gaussian Process: Learning Interpretable Shared Hidden Structure Across Data Spaces for Design Space Analysis and Exploration. *J. Mech. Des* **2020**, *142* (8). https://doi.org/10.1115/1.4046074.

(92)  Bostanabad, R.; Chan, Y.-C.; Wang, L.; Zhu, P.; Chen, W. Globally Approximate Gaussian Processes for Big Data With Application to Data-Driven Metamaterials Design. *J. Mech. Des* **2019**, *141* (11). https://doi.org/10.1115/1.4044257.

(93)  Erickson, C. B.; Ankenman, B. E.; Sanchez, S. M. Comparison of Gaussian Process Modeling Software. *European Journal of Operational Research* **2018**, *266* (1), 179–192. https://doi.org/10.1016/j.ejor.2017.10.002.

(94)  Bastos, L. S.; O'Hagan, A. Diagnostics for Gaussian Process Emulators. *Technometrics* **2009**, *51* (4), 425–438. https://doi.org/10.1198/TECH.2009.08019.

(95)  Chen, H.; Loeppky, J. L.; Sacks, J.; Welch, W. J. Analysis Methods for Computer Experiments: How to Assess and What Counts? *Statistical Science* **2016**, *31* (1), 40–60. https://doi.org/10.1214/15-STS531.

(96)  Kushner, H. J. A New Method of Locating the Maximum Point of an Arbitrary Multipeak Curve in the Presence of Noise. *J. Basic Eng* **1964**, *86* (1), 97–106. https://doi.org/10.1115/1.3653121.

(97)   Jones, D. R. A Taxonomy of Global Optimization Methods Based on Response Surfaces. *Journal of Global Optimization* **2001**, *21* (4), 345–383. https://doi.org/10.1023/A:1012771025575.

(98)   Cox, D. D.; John, S. A Statistical Method for Global Optimization. In *[Proceedings] 1992 IEEE International Conference on Systems, Man, and Cybernetics*; 1992; pp 1241–1246 vol.2. https://doi.org/10.1109/ICSMC.1992.271617.

(99)   Emmerich, M. T. M.; Giannakoglou, K. C.; Naujoks, B. Single- and Multiobjective Evolutionary Optimization Assisted by Gaussian Random Field Metamodels. *IEEE Transactions on Evolutionary Computation* **2006**, *10* (4), 421–439. https://doi.org/10.1109/TEVC.2005.859463.

(100)  Abdolshah, M.; Shilton, A.; Rana, S.; Gupta, S.; Venkatesh, S. Expected Hypervolume Improvement with Constraints. In *2018 24th International Conference on Pattern Recognition (ICPR)*; 2018; pp 3238–3243. https://doi.org/10.1109/ICPR.2018.8545387.

(101)  Yang, K.; Emmerich, M.; Deutz, A.; Bäck, T. Multi-Objective Bayesian Global Optimization Using Expected Hypervolume Improvement Gradient. *Swarm and Evolutionary Computation* **2019**, *44*, 945–956. https://doi.org/10.1016/j.swevo.2018.10.007.

(102)  Wang, Z.; Jegelka, S. Max-Value Entropy Search for Efficient Bayesian Optimization. *arXiv:1703.01968 [cs, math, stat]* **2018**.

(103)  Hernández-Lobato, D.; Hernández-Lobato, J. M.; Shah, A.; Adams, R. P. Predictive Entropy Search for Multi-Objective Bayesian Optimization. *arXiv:1511.05467 [stat]* **2016**.

(104)  Abdolshah, M.; Shilton, A.; Rana, S.; Gupta, S.; Venkatesh, S. Multi-Objective Bayesian Optimisation with Preferences over Objectives. *arXiv:1902.04228 [cs, stat]* **2019**.

(105)  Tran, A.; Eldred, M.; McCann, S.; Wang, Y. SrMO-BO-3GP: A Sequential Regularized Multi-Objective Constrained Bayesian Optimization for Design Applications; American Society of Mechanical Engineers Digital Collection, 2020. https://doi.org/10.1115/DETC2020-22184.

(106)  Biswas, A.; Fuentes, C.; Hoyle, C. A MO-BAYESIAN OPTIMIZATION APPROACH USING THE WEIGHTED TCHEBYCHEFF METHOD. *Journal of Mechanical Design* **2021**, 1–30. https://doi.org/10.1115/1.4051787.

(107)  Biswas, A. Hybrid Statistical and Engineering Optimization Architectures in Early Multidisciplinary Designs of Resilience and Expensive Black-box Complex Systems.

(108)  Knowles, J. ParEGO: A Hybrid Algorithm with on-Line Landscape Approximation for Expensive Multiobjective Optimization Problems. *IEEE Transactions on Evolutionary Computation* **2006**, *10* (1), 50–66. https://doi.org/10.1109/TEVC.2005.851274.

(109)  Daulton, S.; Balandat, M.; Bakshy, E. Differentiable Expected Hypervolume Improvement for Parallel Multi-Objective Bayesian Optimization. *arXiv:2006.05078 [cs, math, stat]* **2020**.

(110)  Blank, J.; Deb, K. Pymoo: Multi-Objective Optimization in Python. *IEEE Access* **2020**, *8*, 89497–89509. https://doi.org/10.1109/ACCESS.2020.2990567.

(111)  Pohlheim, H. Examples of Objective Functions (GEATbx.Com). 2006.

Supplement materials of the paper "*Multi-objective Bayesian optimization of ferroelectric materials with interfacial control for memory and energy storage applications*"

**Appendix:**

### A. Test Problems:

*Test Problem 1: ZDT1*

$$\max_{x_1, x_2} f_1 = \max_{x_1, x_2} x_1 \tag{14a}$$

$$\max_{x_1, x_2} f_2 = \max_{x_1, x_2} \left(1 - \sqrt{\frac{f_1}{g}}\right) \tag{14b}$$

where, $g = 1 + 9x_2$; $0 \leq x_1, x_2 < 1$

*Test Problem 2: 6HC-IAP*

*Maximize Objective1: 2-D Six-hump camel back function:*

$$\max_{x_1, x_2} f_1 = \max_{x_1, x_2} \left(4 - 2.1x_1^2 + \frac{x_1^4}{3}\right)x_1^2 + x_1 x_2 + (-4 + 4x_2^2)x_2^2 \tag{15a}$$

*Maximize Objective2: 2-D Inversed-Ackley's Path function:*

$$\max_{x_1, x_2} f_2 = \max_{x_1, x_2} a \times \exp\left(-b\sqrt{\frac{x_1^2 + x_2^2}{2}}\right) + \exp\left(\frac{\cos(cx_1) + \cos(cx_2)}{2}\right) - a - \exp(1) \tag{15b}$$

where $a = 20$; $b = 0.2$; $c = 2\pi$; $-3 \leq x_1 < 3$; $-2 \leq x_2 < 2$

The values of $a$, $b$ and $c$ are considered as in [111].

## B. *Exhaustive exploration of 2-4-6 KLGD for bulk PZO:*

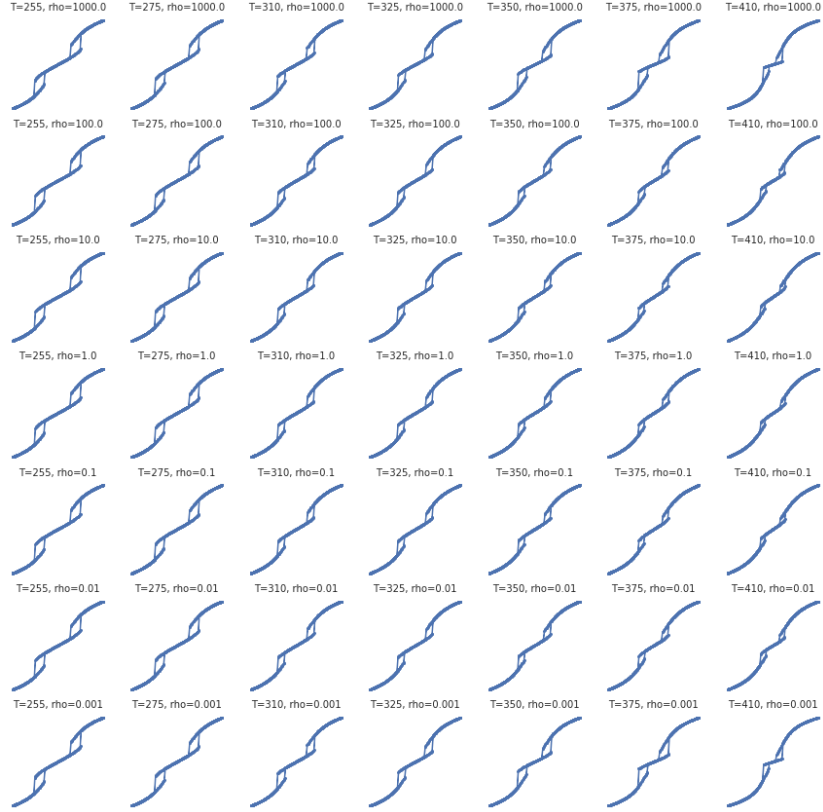

**Figure 13:** *2-4-6 KLGD for bulk PZO:* Loop structures in parameter space $T = [255, 410]K, \rho = [10^{-3}, 10^3], h = 5nm, \Delta_G = 0.2eV$. External field, $E = 3 * \sin(2\omega t)$. This parameter space falls into **AFE.**

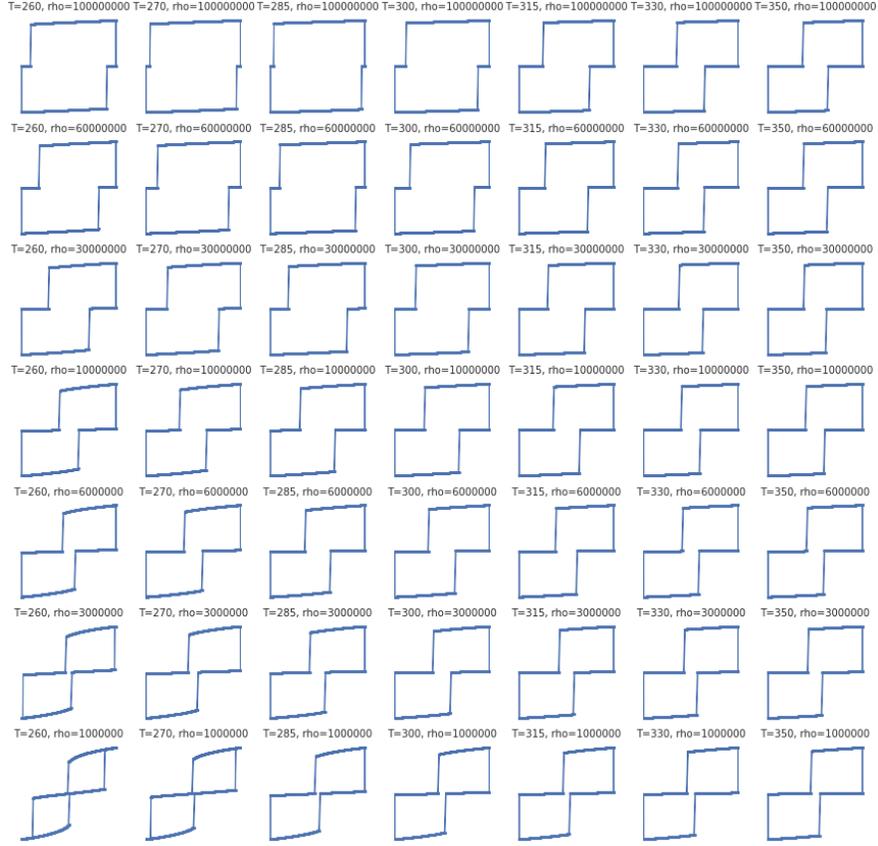

**Figure 14:** *2-4-6 KLGD for bulk PZO*: Loop structures in parameter space $T = [260, 350]K, \rho = [10^6, 10^8], h = 50nm, \Delta_G = 0.2eV$. External field, $E = 3 * \sin(2\omega t)$ . This parameter space falls into **FE.**

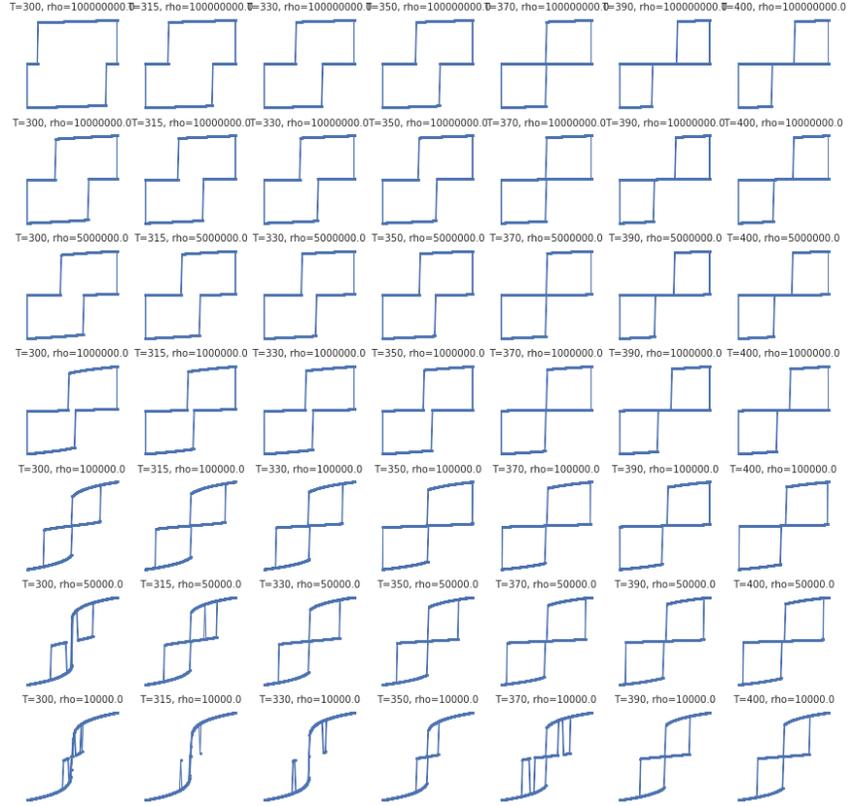

**Figure 15:** *2-4-6 KLGD for bulk PZO*: Loop structures in parameter space $T = [300, 400]K$, $\rho = [10^4, 10^8]$, $h = 50nm$, $\Delta_G = 0.2eV$. External field, $E = 3 * \sin(2\omega t)$ . This parameter space falls into **AFE** and **FE.**

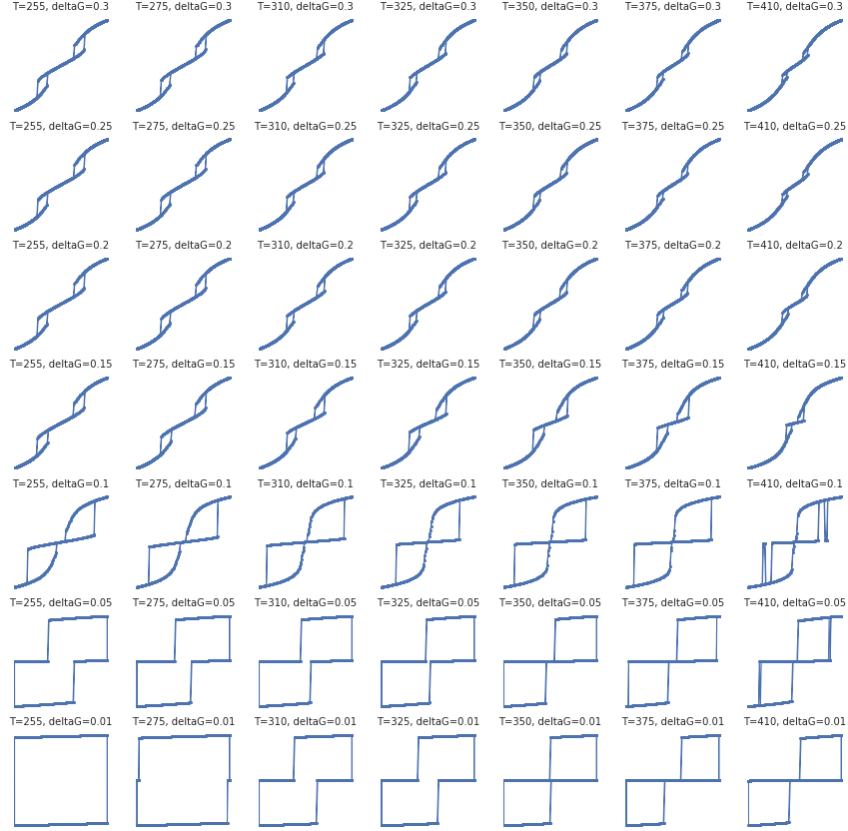

**Figure 16:** *2-4-6 KLGD for bulk PZO*: Loop structures in parameter space $T = [255, 410]K, \rho = 10^2, h = 5nm, \Delta_G = [0.01, 0.3]eV$. External field, $E = 3 * \sin(2\omega t)$. This parameter space falls into **AFE** and **FE**.

The loop structure is configured using downhill simplex algorithm in *scipy* python package (*optimize.fmin*) to find the desired local minima of polarization (P) over the objective of free energy (F), depending on the field history, at each value of field, $E(t)$ in the time-steps. Some of the loops is ill-configured due to the convergence issue of the algorithm (stuck to sub-optima or any undesired local optima region). It is generally a hard problem to choose the optimization algorithm, even harder when we aim for a specific local or global minima. However, it represents the overall picture which we used to compare with the results from MOBO. The parameter exploration done through MOBO, with the defined space of avoiding any adverse effect in building the pareto front due to the stated numerical issue of some loop configurations. As in our analysis, we sometime detect this anomaly once we encounter any outlier (also chooses to be pareto optimal) which caused a sudden drop in the number of pareto solutions and/or the acquisition functions still does not exploit that region. It is to be noted, given a well configured loops, MOBO is working efficiently in terms of accuracy and cost as per the results shown. As the focus of this paper is the implementation of MOBO in physics based model, we ignored this numerical issue in the model itself (loop creation) and assumed the goal is to implement or expand to any application where we have well-configured loops.